\documentclass[sn-mathphys]{sn-jnl}

\jyear{2021}%

\theoremstyle{thmstyleone}%
\theoremstyle{thmstyletwo}%

\theoremstyle{thmstylethree}%

\usepackage{enumitem}
\definecolor{mygray}{gray}{0.}

\usepackage{supertabular}
\usepackage{tabularx}
\usepackage{multirow}

\usepackage[justification=centering]{caption}

\begin{document}

\title[{U-Sleep's resilience to AASM guidelines}]{{U-Sleep's resilience to AASM guidelines}}


\author*[1,2]{\fnm{Luigi} \sur{Fiorillo}}\email{luigi.fiorillo@supsi.ch}
\equalcont{These authors contributed equally to this work.}

\author[1,2]{\fnm{Giuliana} \sur{Monachino}}
\equalcont{These authors contributed equally to this work.}

\author[3]{\fnm{Julia} \sur{van der Meer}}

\author[3]{\fnm{Marco} \sur{Pesce}}

\author[3]{\fnm{Jan D.} \sur{Warncke}}

\author[3]{\fnm{Markus H.} \sur{Schmidt}}

\author[3]{\fnm{Claudio L.A.} \sur{Bassetti}}

\author[1,3]{\fnm{Athina} \sur{Tzovara}}

\author[1]{\fnm{Paolo} \sur{Favaro}}

\author[2]{\fnm{Francesca D.} \sur{Faraci}}

\affil*[1]{\orgdiv{Institute of Informatics}, \orgname{University of Bern}, \orgaddress{\street{Neubr{\"u}ckstrasse 10}, \city{Bern}, \postcode{3012}, \country{Switzerland}}}

\affil[2]{\orgdiv{Institute of Digital Technologies for Personalized Healthcare $\vert$ MeDiTech}, Department of Innovative Technologies, \orgname{University of Applied Sciences and Arts of Southern Switzerland}, \orgaddress{\street{Via la Santa 1}, \city{Lugano}, \postcode{6962}, \country{Switzerland}}}

\affil[3]{\orgdiv{Sleep Wake Epilepsy Center $\vert$ NeuroTec},  Department of Neurology, \orgname{Inselspital, Bern University Hospital, University of Bern}, \orgaddress{\street{Freiburgstrasse 16}, \city{Bern}, \postcode{3010}, \country{Switzerland}}}


\abstract{
AASM guidelines are the result of decades of efforts aiming at standardizing sleep scoring procedure, with the final goal of sharing a worldwide common methodology. The guidelines cover several aspects from the technical/digital specifications, \textit{e.g.}, recommended EEG derivations, to detailed sleep scoring rules accordingly to age. { Automated sleep scoring systems have always largely exploited the standards as fundamental guidelines. {In this context, deep learning has demonstrated better performance compared to classical machine learning}. Our present work shows} that a deep learning based sleep scoring algorithm may not need to fully exploit the clinical knowledge or to strictly adhere to the AASM guidelines. Specifically, we demonstrate that U-Sleep, a state-of-the-art sleep scoring algorithm, can be strong enough to solve the scoring task even using clinically non-recommended or non-conventional derivations, and with no need to exploit information about the chronological age of the subjects. We finally strengthen a well-known finding that using data from multiple data centers always results in a better performing model compared with training on a single cohort. Indeed, we show that this latter statement is still valid even by increasing the size and the heterogeneity of the single data cohort. In all our experiments we used 28528 polysomnography studies from 13 different clinical studies.
}




\maketitle

\section*{Introduction}

Since its origin in the late 1950s, polysomnography (PSG) has been at the centre of sleep medicine testing with the main aim of standardizing and of simplifying the scoring procedure. A common methodology has fostered clinical research and improved sleep disorder classification and comprehension. A PSG typically involves a whole night recording of bio-signals. Brain activity, eye movements, muscle activity, body position, heart rhythm, breathing functions and other vital parameters are monitored overnight. PSG scoring is the procedure of extracting information from the recorded signals. Sleep stages, arousals, respiratory events, movements and cardiac events have to be correctly identified. Wakefulness and sleep stages, \textit{i.e.}, stages 1, 2, 3 and rapid eye movement (REM), can be mainly described by three bio-signals: electroencephalography (EEG), electrooculography (EOG) and electromyography (EMG).
Clinical sleep scoring involves a visual analysis of overnight PSG by a human expert and may require up to two hours of tedious repetitive work. The scoring is done worldwide accordingly to official standards, \textit{e.g.}, the American Academy of Sleep Medicine (AASM) scoring manual {\cite{berry2012rules}}.\\

Artificial intelligence (AI) is a powerful technique that has the potential to simplify and accelerate the sleep scoring procedure. In literature over the last two decades, a wide variety of machine learning (ML) and deep learning (DL) based algorithms have been proposed to solve the sleep scoring task \cite{ronzhina2012sleep, csen2014comparative, radha2014comparison, aboalayon2016sleep, boostani2017comparative, fiorillo2019automated}.
DL based scoring algorithms have shown higher performances compared to the traditional ML approaches.
Autoencoders \cite{tsinalis2016automatic}, deep neural networks (DNNs) \cite{dong2018mixed}, U-Net inspired architectures \cite{perslev2019u, perslev2021u}, convolutional neural networks (CNNs) and fully-CNNs \cite{tsinalis2016automaticCNN, vilamala2017deep, zhang2018complex, chambon2018deep, cui2018automatic, olesen2018deep, patanaik2018end, sors2018convolutional, yildirim2019deep, fiorillo2020temporal}, recurrent neural networks (RNNs) \cite{michielli2019cascaded, phan2019seqsleepnet} and several combinations of them \cite{supratak2017deepsleepnet, biswal2018expert, malafeev2018automatic, stephansen2018neural, back2019intra, mousavi2019sleepeegnet, seo2020intra, phan2020towards, supratak2020tinysleepnet, phan2021xsleepnet} have been recently proposed in sleep scoring. The possibility to extract complex information from a large amount of data is one of the main reasons to apply DL techniques in PSG classification. Another significant advantage is the ability to learn features directly from raw data, by also taking into account the temporal dependency among the sleep stages.\\

In literature we can find many examples about how clinical guidelines have been exploited when trying to support ML and DL based algorithms. The oldest Rechtschaffen and Kales (R\&K) \cite{rechtschaffen1968manual} or the updated AASM {\cite{berry2012rules}} scoring manuals have been designed to cover all the aspects of the PSG: from the technical/digital specifications (\textit{e.g.}, assessment protocols, data filtering, recommended EEG derivations) to the scoring rules (\textit{e.g.}, sleep scoring rules for adults, children and infants, movement rules, respiratory rules) and the final interpretation of the results. {All the sleep scoring algorithms, both ML or DL based, are trained on sleep recordings annotated by sleep physicians according to these manuals. In some of these studies the sleep recordings are pre-filtered, as indicated in the AASM guidelines, before feeding them to their scoring system. Almost all of the algorithms mentioned above are trained using recommended channel derivations and fixed length (\textit{i.e.}, 30-second) sleep epochs.} However, it still remains unknown whether a DL based sleep scoring algorithm actually needs to be trained by following these guidelines. More than a decade ago, it was already highlighted that sleep is not just a global phenomenon affecting the whole brain at the same time, but that sleep patterns such as slow waves and spindle oscillations often occur out-of-phase in different brain regions \cite{huber2004local}. 
Hence, it may be that DL based scoring algorithms could retrieve the needed information from brain regions that are not necessarily the ones indicated in the AASM guidelines, reaching equally high performance.
Indeed, in the growing field of mobile sleep monitoring with wearable devices, many studies are attempting to tackle the automated sleep scoring task by using unconventional channels, {even not necessarily placed on the scalp, \textit{e.g.}, in-ear EEG \cite{nakamura2019hearables, mikkelsen2021sleep, jorgensen2020ear}}. Furthermore, in the AASM manual and in previous studies \cite{Ohayon2004, kocevska2021sleep}, age has been addressed as one of the demographic factors that mainly change sleep characteristics (\textit{e.g.}, sleep latency, sleep cycle structure, EEG amplitude etc.). To the best of our knowledge, it has never been attempted before to incorporate this information within a sleep scoring system: it could reasonably improve its performance.\\

To date, all the efforts have focused on optimizing a sleep scoring algorithm in order to be ready to score any kind of subject. {Data heterogeneity is one of the biggest challenges to address. A common objective among researchers is to increase the model generalizability, \textit{i.e.}, the ability of the model to make accurate predictions over different or never seen data domains. The performance of a sleep scoring algorithm on a PSG from an unseen data distribution (\textit{e.g.}, different data domains/centers) usually drastically decreases {\cite{phan2020towards, guillot2021robustsleepnet, perslev2021u, olesen2021automatic, vallat2021open}}. This drop in performance can be due to a variety of well-known reasons: high inter-scorer variability; hardware variability, \textit{e.g.}, channels/derivations; high data variability from different sleep centers, \textit{e.g.}, subject distributions with different sleep disorders.} In recent studies, Phan et al. and Guillot et al. \cite{phan2020towards, guillot2021robustsleepnet} propose to adapt a sleep scoring architecture on a new data domain via transfer learning techniques. They demonstrate the efficiency of their approaches in addressing the variability between the source and target data domains. Perslev at al., Olesen et al. and Vallat et al. \cite{perslev2021u, olesen2021automatic, vallat2021open} propose to train their sleep scoring architectures on tens of thousands of PSGs from different large-scale-heterogeneous cohorts. They demonstrate that using data from many different sleep centers improves the performance of their model, even on never seen data domains. In particular, Olesen et al. \cite{olesen2021automatic} show that models trained on a single data domain fail to generalize on a new data domain or data center. \\

In our study we {do} several experiments to evaluate the resilience of an existing DL based algorithm against the AASM guidelines. In particular we {focus} on the following questions:\\

\begin{enumerate}[label=(\roman*),font=\itshape]

\item can a sleep scoring algorithm successfully encode sleep patterns, from clinically non-recommended or non-conventional electrode derivations?\\

\item can a single sleep center large dataset contain enough heterogeneity (\textit{i.e.}, different demographic groups, different sleep disorders) to allow the algorithm to generalize on multiple data centers?\\

\item whenever we train an algorithm on a dataset with subjects with a large age range, should we exploit the information about their age, conditioning the training of the model on it?\\

\end{enumerate}

We {run} all of our experiments on U-Sleep, a state-of-the-art sleep scoring architecture recently proposed by Perslev et al. \cite{perslev2021u}. U-Sleep has been chosen mainly for the following reasons: it has been evaluated on recordings from 15660 participants of 16 different clinical studies (four of them never seen by the architecture); it processes inputs of arbitrary length, from any arbitrary EEG and EOG electrode positions, from any hardware and software ﬁltering; it predicts the sleep stages for an entire PSG recording in a single forward pass; it outputs sleep stage labels at any temporal frequency, up to the signal sampling rate, \textit{i.e.}, it can label sleep stages at shorter intervals than the standard 30-seconds, up to one sleep stage per each sampled time point.\\

In the original implementation of U-Sleep we found an extremely interesting \textit{bug}: the data sampling procedure was not extracting the channel derivations recommended in the AASM guidelines, as stated by the authors in \cite{perslev2021u}. Instead, atypical or non-conventional channel derivations were randomly extracted. This insight triggered the above mentioned question \textit{(i)}. \\

Our contributions can be summarized as follows: (1) we find that a DL sleep scoring algorithm is still able to solve the scoring task, with high performance, even when trained with clinically non-conventional channel derivations; (2) we show that a DL sleep scoring model, even if trained on a single large and heterogeneous sleep center, fails to generalize on new recordings from different data centers; (3) {we show that the conditional training based on the chronological age of the subjects does not improve the performance of a DL sleep scoring architecture}. 

\section*{Results}

\subsection*{Datasets and model experiments}

We train and evaluate U-Sleep on 19578 recordings from 15322 subjects of 12 publicly available clinical studies, as done previously \cite{perslev2021u}. \\

In this study we also exploit the Bern Sleep Data Base (BSDB) registry, the sleep disorder patient cohort of the Inselspital, University hospital Bern. The recordings have been collected from 2000 to 2021 at the Department of Neurology, at the University hospital Bern. Secondary usage was approved by the cantonal ethics  committee (KEK-Nr. 2020-01094). The dataset consists of 8950 recordings from patients and healthy subjects aged 0-91 years. {In our experiments we consider 8884 recordings, given the low signal quality of the remaining recordings.} The strength of this dataset is that, unlike the ones available online, it contains patients covering the full spectrum of sleep disorders, many of whom were diagnosed with multiple sleep disorders and non-sleep related comorbidities \cite{mathis2022u}; thus providing an exceptionally heterogeneous PSG data set.\\

{An overview of the BSDB and the open access (OA) datasets along with demographic statistics is reported in Table~\ref{db_overview}. In Supplementary notes: Datasets, we also report a detailed description of all the datasets used in this study.} \\

\begin{table}[t]
\caption{}
\label{db_overview}
\begin{center}
\begin{tabular}{lllll}
Datasets & Recordings & Age (years) & Sex \% (F/M)\\
\hline
\noalign{\vskip 1mm}
\cite{zhang2018national, bakker2018gastric} ABC (\checkmark) & 132 & 48.8 $\pm$ 9.8 & 43/57 \\
\hline
\noalign{\vskip 1mm}
\cite{zhang2018national, rosen2003prevalence} CCSHS (\checkmark) & 515 & 17.7 $\pm$ 0.4 & 50/50 \\
\hline
\noalign{\vskip 1mm}
\cite{zhang2018national, redline1995familial} CFS (\checkmark) & 730 & 41.7 $\pm$ 20.0 & 55/45 \\
\hline
\noalign{\vskip 1mm}
\cite{zhang2018national, marcus2013randomized, redline2011childhood} CHAT (\checkmark) & 1638 & 6.6 $\pm$ 1.4 & 52/48 \\
\hline
\noalign{\vskip 1mm}
{\cite{perslev2021u}} DCSM \checkmark& 255 & - & - \\
\hline
\noalign{\vskip 1mm}
\cite{zhang2018national, rosen2012multisite} HPAP (\checkmark) & 238 & 46.5 $\pm$ 11.9 & 43/57 \\
\hline
\noalign{\vskip 1mm}
\cite{zhang2018national, chen2015racial} MESA (\checkmark) & 2056 & 69.4 $\pm$ 9.1 & 54/46 \\
\hline
\noalign{\vskip 1mm}
\cite{zhang2018national, blackwell2011associations, osteoporotic2015relationships} MROS (\checkmark) & 3926 & 76.4 $\pm$ 5.5 & 0/100 \\
\hline
\noalign{\vskip 1mm}
\cite{goldberger2000physiobank, ghassemi2018you} PHYS \checkmark & 994 & 55.2 $\pm$ 14.3 & 33/67 \\
\hline
\noalign{\vskip 1mm}
\cite{goldberger2000physiobank, kemp2000analysis} SEDF-SC \checkmark & 153 & 58.8 $\pm$ 22.0 & 53/47 \\
\hline
\noalign{\vskip 1mm}
\cite{goldberger2000physiobank, kemp2000analysis} SEDF-ST \checkmark & 44 & 40.2 $\pm$ 17.7 & 68/32 \\
\hline
\noalign{\vskip 1mm}
\cite{zhang2018national, quan1997sleep} SHHS (\checkmark) & 8444 & 63.1 $\pm$ 11.2 & 52/48 \\
\hline
\noalign{\vskip 1mm}
\cite{zhang2018national, cummings1990appendicular, spira2008sleep} SOF (\checkmark) & 453 & 82.8 $\pm$ 3.1 & 100/0 \\
\hline
\hline
\noalign{\vskip 1mm}
BSDB & 8884 & 47.9 $\pm$ 18.4 & 66/34 \\
\hline
\end{tabular}
\end{center}
\end{table}

The data pre-processing and the data selection/sampling across all the datasets is implemented as described in \cite{perslev2021u} (see subsection U-Sleep architecture). In contrast with the recommendation of the AASM manual, no filtering was applied to the EEG and the EOG signals during the pre-processing procedure. {Most importantly, we found that in the original implementation of U-Sleep \cite{perslev2021u} atypical or non-conventional channel derivations were erroneously extracted.} In fact, the data extraction and the resulting sampling procedure were creating totally random derivations, see Supplementary Table 6, obviously different to those recommended in the AASM guidelines.
In this study, we examine the resilience of U-Sleep with respect to the official AASM gruidelines. To this aim, we extract the channel derivations following the guidelines (as was originally meant to be done in \cite{perslev2021u}), to better understand the impact of channel selection on the overall performance. Below we summarize all the experiments performed in our work on U-Sleep:\\

\begin{enumerate}[label=(\roman*),font=\itshape]
\item We pre-train U-Sleep on all the OA datasets using both the original implementation selecting the atypical channel derivations (U-Sleep-v0), and our adaptation following AASM guidelines (U-Sleep-v1). We split each dataset in training (75\%), validation (up to 10\%, at most 50 subjects) and test set (up to 15\%, at most 100 subjects). The split of the PSG recordings is done per-subject or per-family, \textit{i.e.}, recordings from the same subject or members of the same family appear in the same data split. In Supplementary Table 7 we summarize the data split on each OA dataset. We evaluate both U-Sleep-v0 and U-Sleep-v1 on the test set of the BSDB dataset. We also evaluate the models on the whole BSDB\textsubscript{(100\%)} dataset, to test on a higher number of subjects, with a higher heterogeneity of sleep disorders and a wider age range. A model pre-trained on the OA datasets and evaluated directly on the BSDB dataset is what we will refer to as direct transfer (DT) on BSDB.\\
\item We exploit the BSDB dataset to evaluate whether a DL based scoring architecture, trained with a large and a highly heterogeneous database, is able to generalize on the OA datasets from different data centers. We split the BSDB recordings in training (75\%), validation (10\%) and test set (15\%). We run two different experiments on U-Sleep-v1: we train the model from scratch (S) on the BSDB dataset; we fine-tune (FT) the model pre-trained in \textit{(i)} on the BSDB dataset, by using the transfer learning approach (see subsection Transfer learning). Then, we evaluate both (S) and (FT) on the test set of all the OA datasets and the test set of the BSDB dataset.\\
\item  We exploit the BSDB dataset to investigate whether U-Sleep needs to be trained by also having access to chronological age-related information. We split the BSDB dataset in seven groups, according to the age categories of the subjects \cite{Ohayon2004}, resulting in $G=7$ sub-datasets, see Supplementary notes: Age analysis. We further split the recordings of each subdataset in training (75\%), validation (10\% at most 50 subjects) and test set (15\% at most 100 subjects). We run three different experiments on U-Sleep-v1: we fine-tune the model by using all the training sets of the seven groups (FT); we fine-tune seven independent models by using the training set of each group independently (FT-I); we fine-tune a single sandwich batch normalization model (exploiting the batch normalization layers, see subsection Conditional learning), to add the condition on the age-group-index $G$ for each recording (FT-SaBN). These last two experiments are replicated considering only two age groups, \textit{i.e.}, babies/children and adults, as recommended in {\cite{berry2012rules}}, resulting in two additional fine-tuned model (FT-I and FT-SaBN for $G=2$). We then evaluate all of the fine-tuned models on the independent test set of each age group.\\
\end{enumerate}

In Supplementary Table 8 we summarize the two different data split sets, in experiment \textit{(ii)} and experiment \textit{(iii)}, on the BSDB dataset.\\

\subsection*{Performance overview}

\begin{enumerate}[label=(\roman*),font=\itshape]
\item \textit{Clinically non-recommended channel derivations}.
In Table~\ref{usleep_i} we compare the performance of U-Sleep pre-trained on all the OA datasets, with (U-Sleep-v0) and without (U-Sleep-v1) using randomly ordered channel derivations. There is no statistically significant difference between the two differently trained architectures evaluated on the test set of the BSDB dataset ({two-sided} paired t-test $p-value > 0.05$). Most importantly, we find no difference in performance with the direct transfer also on the whole BSDB\textsubscript{(100\%)} dataset ({two-sided} paired t-test $p-value > 0.05$). 
These results clearly show how the architecture is able to generalize regardless of the channel derivations used during the training procedure, also on a never seen highly heterogeneous dataset. In Supplementary Table 9 we also compare the performance of U-Sleep-v0 and U-Sleep-v1 per sleep stage. {The results suggest that there are statistically significant differences between the two differently trained architectures for each of the classes (two-sided paired t-test $p-value < 0.001$). U-Sleep-v0 better recognizes N1 and N3 sleep stages, at the expense of awake, N2 and REM sleep stages.}\\

\begin{table}[t]
\caption{}
\label{usleep_i}
\begin{center}
\begin{tabular}{lll}
Datasets & U-Sleep-v0 & U-Sleep-v1\\
\hline
\noalign{\vskip 1mm}
BSDB &  72.5 $\pm$ 12.2 & 72.5 $\pm$ 12.0 \\
\hline
\noalign{\vskip 1mm}
BSDB\textsubscript{(100\%)} & 72.9 $\pm$ 12.4 & 72.9 $\pm$ 12.4 \\
\hline
\end{tabular}
\end{center}
\end{table}

\begin{table}[t]
\caption{}
\label{usleep_ii}
\begin{center}
\begin{tabular}{llll}
Datasets & U-Sleep-v1 & U-Sleep-v1 (S) & U-Sleep-v1 (FT)\\
\hline
\noalign{\vskip 1mm}
ABC & 73.6 $\pm$ 11.4 & 71.4 $\pm$ 13.9 & 69.0 $\pm$ 12.5\\
\hline
\noalign{\vskip 1mm}
CCSHS & 84.9 $\pm$ 5.1 & 77.3 $\pm$ 7.2 & 77.3 $\pm$ 6.7\\
\hline
\noalign{\vskip 1mm}
CFS & 76.6 $\pm$ 11.6 & 70.2 $\pm$ 10.8 & 70.9 $\pm$ 10.2\\
\hline
\noalign{\vskip 1mm}
CHAT & 82.1 $\pm$ 6.5 & 72.9 $\pm$ 8.0 & 68.8 $\pm$ 8.7\\
\hline
\noalign{\vskip 1mm}
DCSM & 79.3 $\pm$ 9.3 & 71.5 $\pm$ 11.2 & 69.3 $\pm$ 10.5\\
\hline
\noalign{\vskip 1mm}
HPAP & 73.8 $\pm$ 10.8 & 68.9 $\pm$ 11.1 & 67.9 $\pm$ 12.5\\
\hline
\noalign{\vskip 1mm}
MESA & 72.7 $\pm$ 10.8 & 68.5 $\pm$ 14.3 & 68.7 $\pm$ 11.9\\
\hline
\noalign{\vskip 1mm}
MROS & 71.4 $\pm$ 12.1 & 61.7 $\pm$ 13.7 & 63.9 $\pm$ 13.2\\
\hline
\noalign{\vskip 1mm}
PHYS & 74.2 $\pm$ 10.7 & 72.9 $\pm$ 11.2 & 73.2 $\pm$ 11.4\\
\hline
\noalign{\vskip 1mm}
SEDF-SC & 77.8 $\pm$ 7.9 & 75.8 $\pm$ 8.0 & 77.9 $\pm$ 7.7\\
\hline
\noalign{\vskip 1mm}
SEDF-ST & 77.2 $\pm$ 10.1 & 64.3 $\pm$ 15.4 & 67.5 $\pm$ 12.4\\
\hline
\noalign{\vskip 1mm}
SHHS & 76.9 $\pm$ 9.7 & 70.9 $\pm$ 9.3 & 73.0 $\pm$ 8.9\\
\hline
\noalign{\vskip 1mm}
SOF & 74.8 $\pm$ 9.8 & 64.6 $\pm$ 12.6 & 67.5 $\pm$ 11.2\\
\hline
\noalign{\vskip 1mm}
avg OA & 76.5 $\pm$ 10.6 & 69.9 $\pm$ 11.9 & 70.2 $\pm$ 11.1\\
\hline
\hline
\noalign{\vskip 1mm}
BSDB & 72.5 $\pm$ 12.0 \textsuperscript{(DT)} & 77.6 $\pm$ 11.3 & 77.3 $\pm$ 11.4\\
\hline
\end{tabular}
\end{center}
\end{table}

\item \textit{Generalizability on different data centers with a heterogeneous dataset}.\\
In Table~\ref{usleep_ii} we report the results obtained on U-Sleep-v1 pre-trained \textit{(i)} on the OA datasets, and evaluated on all the test sets of the OA datasets and on the test set of the BSDB dataset. We also show the results obtained on U-Sleep-v1 trained from scratch (S) on the BSDB dataset, and the results obtained on the model pre-trained in \textit{(i)} on OA and then fine-tuned (FT) on the BSDB dataset. Unlike what we expected, both the models (S) and (FT), trained with a large and a highly heterogeneous database, are not able to generalize on the OA datasets from the different data centers. The average performance achieved on the OA with (S) and (FT) models is significantly lower compared to the performance of the model pre-trained on OA ({two-sided} paired t-tests $p-value < 0.001$). Whilst, with both (S) and (FT) we show a significant increase in performance compared to the direct transfer (DT), on the test set of the BSDB dataset ({two-sided} paired t-tests $p-value < 0.001$). We also find that the training from scratch results in significantly higher performance ({two-sided} paired t-test $p-value < 0.001$) on the BSDB dataset, compared to the performance of the fine-tuned model. No significant difference ({two-sided} paired t-test $p-value > 0.05$) occurs between (S) and (FT) evaluated on the average performance on OA datasets.
The pre-training on the OA dataset is not beneficial for the model fine-tuned on the BSDB dataset. With a large number of highly heterogeneous subjects, we can directly train the model from scratch on the dataset. However, we have to mention that the main advantage of using the fine-tuned model is that it reaches same performance in less computational time, \textit{i.e.}, a fewer number of iterations (number of iterations: FT $=382 <$ S $=533$). \\

\begin{table}[t]
\caption{}
\label{usleep_iii}
\begin{center}
\resizebox{\textwidth}{!}{\begin{tabular}{llllll}
Age groups & FT-G1 & FT-I-G7 & FT-I-G2 & FT-SaBN-G7 & FT-SaBN-G2\\
 \hline
\noalign{\vskip 1mm}
B & 74.9 $\pm$ 6.8 & 74.1 $\pm$ 6.6 \textsuperscript{{G\textsubscript{1}}} & 74.8 $\pm$ 6.2 \textsuperscript{{G\textsubscript{1}}} & 72.2 $\pm$ 7.7 & 72.6 $\pm$ 7.7 \\
\hline
\noalign{\vskip 1mm}
C & 75.0 $\pm$ 9.8 & 74.9 $\pm$ 9.2 \textsuperscript{{G\textsubscript{2}}} & 75.9 $\pm$ 9.1 \textsuperscript{{G\textsubscript{1}}} & 74.8 $\pm$ 8.9 & 75.6 $\pm$ 10.1 \\
\hline
\noalign{\vskip 1mm}
A & 82.7 $\pm$ 13.7 & 80.0 $\pm$ 14.6 \textsuperscript{{G\textsubscript{3}}} & 82.8 $\pm$ 13.6 \textsuperscript{{G\textsubscript{2}}} & 82.3 $\pm$ 13.7 & 82.0 $\pm$ 14.0 \\
\hline
\noalign{\vskip 1mm}
YA & 80.8 $\pm$ 11.5 & 80.6 $\pm$ 11.6 \textsuperscript{{G\textsubscript{4}}} & 80.6 $\pm$ 11.6 \textsuperscript{{G\textsubscript{2}}} & 80.3 $\pm$ 11.9 & 79.9 $\pm$ 11.9 \\
\hline
\noalign{\vskip 1mm}
MA & 80.4 $\pm$ 7.8 & 79.90 $\pm$ 8.0 \textsuperscript{{G\textsubscript{5}}} & 79.8 $\pm$ 8.2 \textsuperscript{{G\textsubscript{2}}} & 79.6 $\pm$ 8.0 & 79.4 $\pm$ 8.3 \\
\hline
\noalign{\vskip 1mm}
E & 75.7 $\pm$ 10.1 & 74.2 $\pm$ 10.7 \textsuperscript{{G\textsubscript{6}}} & 74.9 $\pm$ 10.2 \textsuperscript{{G\textsubscript{2}}} & 74.5 $\pm$ 10.6 & 73.9 $\pm$ 10.9 \\
\hline
\noalign{\vskip 1mm}
OE & 75.2 $\pm$ 11.7 & 73.9 $\pm$ 11.0 \textsuperscript{{G\textsubscript{7}}} & 74.9 $\pm$ 11.3 \textsuperscript{{G\textsubscript{2}}} & 73.8 $\pm$ 11.7 & 74.0 $\pm$ 11.3 \\
\hline
\noalign{\vskip 1mm}
avg & 77.9 $\pm$ 10.7 & 77.0 $\pm$ 10.8 & 77.6 $\pm$ 10.7 & 76.9 $\pm$ 11.0 & 76.8 $\pm$ 11.1 \\
\hline
\end{tabular}}
\end{center}

\end{table}

\item \textit{Training conditioned by age}. In Table~\ref{usleep_iii} we first show the performance of U-Sleep-v1 fine-tuned on all the training sets of the seven BSDB groups, \textit{i.e.}, single model {(FT-G1)}. We also report the performance achieved using the training set of each group independently (FT-I) with $G=7$ and $G=2$ respectively (\textit{i.e.}, seven and two models), and the performance achieved using the training set of the seven/two BSDB groups conditioned (FT-SaBN) by $G=7$ and by $G=2$ groups respectively (\textit{i.e.}, single model). The mean and the standard deviation of the F1-score (\%F1), are computed across the recordings of the test set of each of the seven BSDB age groups. 
Comparing both the experiments (FT-I and FT-SaBN) and types of grouping ($G=2$ and $G=7$) with the baseline (FT), we do not find a statistically significant increase of the performance in any of the subgroups ({one-sided} paired t-test $p-value > 0.05$).
Despite the lack of significant performance differences in our age-conditioned models, REM sleep seems to be less accurately predicted for small children, if the training data set only consists of data from adults (see Supplementary Figure 13, confusion matrix for test $\{CH\}$ against Model 1b). This is an interesting finding since small children exhibit more REM sleep (see Supplementary Figure 11). Visual scoring guidelines for small children differ from the guidelines for adults, with REM sleep scoring strongly relying on irregular respiration \cite{grigg2016visual}.
However, overall these results show that, despite the age-related differences, the DL algorithm is able to deal with different age subgroups at the same time, without needing to have access to chronological age-related information during the training procedure.
\end{enumerate}

\section*{Discussions}

In this paper we demonstrate the resilience of a DL network, when trained on a large and heterogeneous dataset. We focus on the three more significant influencing factors: channel derivation selection, multi center heterogeneity needs and age conditioned fine tuning. Channel derivations do have complementary information, and a DL based model resulted resilient enough to be able to extract sleep patterns also from atypical and clinically non-recommended derivations. We show that the variability among different sleep data centers (\textit{e.g.}, hardware, subjective interpretation of the scoring rules etc.) needs to be taken into account more than the variability inside one single sleep center. A large database such as the BSDB (sleep disorder patient cohort of the Inselspital, with patients covering the full spectrum of sleep disorders) does not have enough heterogeneity to strengthen the performance of the DL based model on unseen data centers. Lastly, we show that a state-of-the-art DL network is able to deal with different age groups simultaneously, mitigating the need of adding chronological age-related information during training. {In summary, what seems to be essential for the visual scoring (\textit{e.g.}, specific channel derivations, or specific scoring rules that consider also the age of the individuals) is not necessary for the DL based automatic procedure, which follows other analysis principles.} \\

The resilience of the DL based model to the atypical or non-conventional channel derivations is fascinating. The model still learns relevant sleep patterns while solving the scoring tasks with high state-of-the-art performance on multiple large-scale-heterogeneous data cohorts. This result proves and strengthens the feasibility to exploit alternative channels to the AASM standard ones(\textit{e.g.}, wearable applications). Although this is a remarkable finding, it would be useful to further investigate the reasons why the DL model is still able to encode clinically valid information. DL has been criticised for its non-interpretability and its black-box behavior, factors that may actually limit its implementation in sleep centers. {Future works, strongly linked to the hot topic of the explainable AI, should focus on solving the following open questions: which sleep patterns/features our DL algorithms are encoding/highlighting from the typical/atypical channel derivations? How each individual channel affects the performance of the DL algorithms?}\\

AASM scoring rules have been widely criticized over the years, for various reasons. The scoring manual has been designed to consider the sleep stages almost as discrete entities. However, it is well-known that sleep should be viewed as a continuum/gradual transition from one stage to another. {A growing consensus suggests that we should reconsider the AASM scoring rules and the entire scoring procedure. Given the high variability among the individual scorers and different sleep centers, more efforts should be made by the scientific community to improve the standardization of the scoring procedure. Perhaps the introduction, even partially, of automated procedure could help.\\
The inter-scorer variability inevitably affects the performance of any kind of algorithm, since all algorithms are learning from the noisy variability of labels. A very relevant finding of this paper is that the heterogeneity given by data coming from different sleep data centers (\textit{e.g.}, different sleep scorers) is much more relevant than the variability coming from patients affected by different sleep disorders.
These latter insights raise a research question yet to be answered: \textit{i.e.}, how could we define and quantify the heterogeneity of a sleep database? To what extent could we consider a database heterogeneous enough, to allow the algorithm to generalize across different data domains/centers?}\\

The age-related findings drive to another important observation: the DL algorithm is intrinsically encoding age-related features, which may not be categorized into discrete age-subgroups. As sleep should be considered as a continuous physiological process, the hyperspace of features associated to the respective age-subgroups should be considered continuum as well. We are forcing the algorithm to learn sleep patterns based on the chronological age of the subjects, but there are many other factors that the DL model is taking into account. Certainly, biological age has an effect on sleep characteristics. Although the DL algorithm does not need to be guided with the chronological age information during its learning procedure, it may be that with a less optimal DL based approach (\textit{e.g.}, architecture, number of channel derivation in input) age would still be a useful information to give in input.\\

To our knowledge, our study on the automatic sleep scoring task is the largest in terms of number of polysomnography recordings and diversity with respect to both patient clinical pathology and age spectrum.\\

{Considering the previous study findings and our present results, the strong resilience and the generalization capability of a DL based architecture is undeniable. DL algorithms are now reaching better performance than the feature based approach. DL is definitely able to extract feature representations that are extremely useful to generalize across datasets from different sleep data centers. These hidden feature representations seem to better decode the unconscious analytical evaluation process of the human scorer. To conclude, being the AASM so widely criticized, being the sleep labels so noisy (\textit{e.g.}, high inter- and intra- scorer variability), and being sleep so complex: could an unsupervised DL based sleep scoring algorithm, that does not need to learn from the labels, be the solution?}

\section*{Methods}

\subsection*{U-Sleep architecture}

U-Sleep \cite{perslev2021u}, optimized version of its predecessor U-Time \cite{perslev2019u}, is inspired by the popular U-Net architecture for image-segmentation \cite{ronneberger2015u, falk2019u, brandt2020unexpectedly}. Below we briefly describe U-Sleep architecture, for further details we refer the reader to \cite{perslev2021u}.\\

U-Sleep is a fully convolutional deep neural network. It takes as input a sequence of length $L$ of 30-second epochs and outputs the predicted sleep stage for each epoch. The peculiarity of this architecture is that it defines the general function $f(\textbf{X};\theta):\mathbb{R}^{L\cdot i\times C}\rightarrow\mathbb{R}^{L\times K}$, where $L>0$ is any positive integer, $\theta$ are the learning parameters, $L$ is a number of fixed-length windows with $i$ sampled points each, $C$ the number of PSG channels and $K$ the number of sleep stages. Hence, U-Sleep takes in input any temporal section of a PSG (even the whole PSG) and output a sequence of labels for each fixed-length $i>0$ window. Ideally $L\cdot i>4096$, because U-Sleep contains 12 pooling operations, downsampling the signal by a factor of 2. The architecture requires at least $C=2$, one EEG and one EOG channel, sampled/resampled at 128Hz, with $K=5$, \textit{i.e.}, ${awake, N1, N2, N3, R}$. \\

\begin{figure}[h]
\centering
\includegraphics[width=0.9\textwidth]{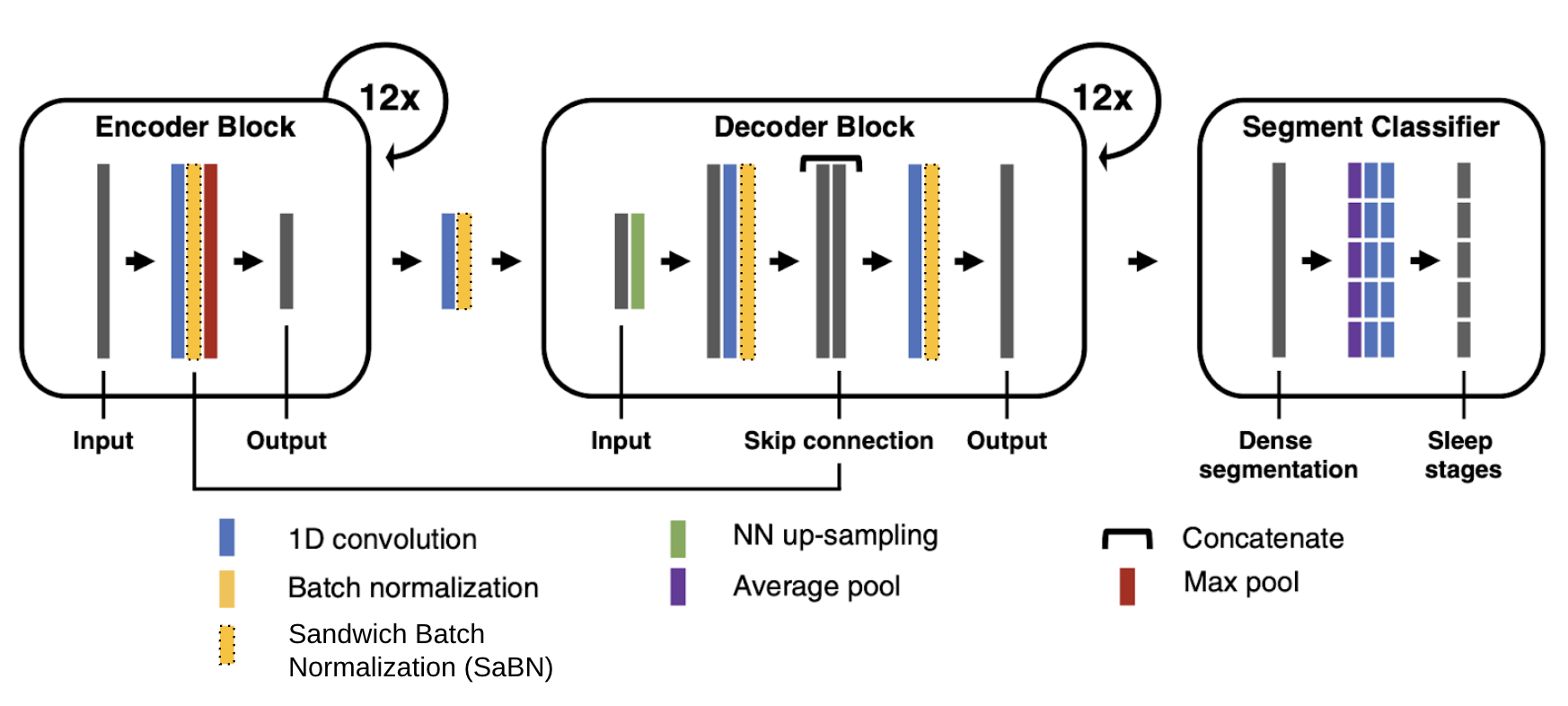}
\caption{}
\label{fig:usleep_architecture}
\end{figure}

U-Sleep architecture consists of three learning modules as shown in Figure~\ref{fig:usleep_architecture}.\\

\begin{itemize}
\item The \textit{encoder} module is designed to extract feature maps from the input signals, each resulting in a lower temporal resolution compared to its input. It includes 12 encoder blocks. Each block consists of a 1D convolutional layer, one layer of activation function - \textit{i.e.}, exponential linear unit (ELU), a batch normalization (BN) layer and one max-pooling layer. 

\item The \textit{decoder} module is designed to up-scale the feature maps to match the temporal resolution of the signals in input. We can interpret the output of the decoder as a high-frequency representation of the sleep stages at the same $f_\textsubscript{s}$ of the input signal (\textit{e.g.}, with $f_\textsubscript{s}=128Hz$, output one sleep stage each $1/128Hz$). The module includes 12 decoder blocks. Each block consists of a nearest neighbour up-sampling layer (\textit{e.g.}, with a $kernel\_size=2$, the length of the feature map in input is doubled), a 1D convolutional layer, one layer of ELU activation function and a BN layer. Then, a skip connection layer combines the up-scaled input with the output of the BN layer of the corresponding encoder block. Finally, a 1D convolution, a ELU non-linearity and a BN are applied to the stacked feature maps. The output has the same temporal resolution of the signal in input. 

\item The \textit{segment classifier} module is designed to segment the high-frequency representation output of the decoder into the desired sleep stage prediction frequency. The module consist of a dense segmentation layer (\textit{i.e.}, 1d convolution layer with a hyperbolic tangent activation function), an average-pooling layer (\textit{e.g.}, with $kernel\_size=stride\_size=30sec*f_\textsubscript{s}$ considering the same prediction frequency of a sleep scorer) and two 1D convolutional layers (the first using an ELU activation function, and the latter using a softmax activation function). The output of the segment classifier is a $L \times K$, where $L$ is the number of segments and $K=5$ is the number of sleep stages.\\
\end{itemize}

The sequence length $L$, the number of filters, the kernel and the stride sizes are specified in Figure~\ref{fig:usleep_architecture}. The softmax function, together with the cross-entropy loss function, is used to train the model to output the probabilities for the five mutually exclusive classes $K$ that correspond to the five sleep stages. The architecture is trained end-to-end via backpropagation, using the sequence-to-sequence learning approach. The model is trained using mini-batch Adam gradient-based optimizer \cite{kingma2014adam} with a learning rate $lr$. The training procedure runs up to a maximum number of iterations, as long as the break early stopping condition is satisfied.\\

Unlike \cite{perslev2021u}, we consider early stopping and data augmentation as regularization techniques. As stated in \cite{goodfellow2016deep} "\textit{regularization is any modiﬁcation we make to a learning algorithm that is intended to reduce its generalization error but not its training error}". Early stopping and data augmentation do so in different ways, they both decrease the regularization error. {By using the early stopping the training procedure is stopped as soon as the performance (\textit{i.e.}, F1-score) on the validation set is lower than it was in the previous iteration steps, by fixing the so called \textit{patience} parameter. By using the data augmentation technique, the signals in input are randomly modified during training procedure to improve model generalization. Variable length of the sequences in input are replaced with a Gaussian noise. For each sample in a batch, with 0.1 probability, a fraction of the sequence is replaced with $N(\mu=\hat{\mu}, \sigma^2=0.01)$, where $\hat{\mu}$ is the mean of the sample's signals. The fraction is sampled with a log-uniform distribution $\{min=0.001; max=0.33\}$. With a 0.1 probability at most one channel is entirely replaced by noise.} \\

The training parameters (\textit{e.g.}, Adam-optimizer parameters beta1 and beta2, mini-batch size etc.) are all set as stated in \cite{perslev2021u}. The learning rate, the early stopping \textit{patience} parameter and the maximum number of iterations have been changed to $10^-5$, $100$ and $1000$ respectively, to let U-Sleep converge faster.
The architecture has several hyperparameters (\textit{e.g.}, number of layers, number/sizes of ﬁlters, regularization parameters, training parameters etc.) which could be optimized to tune its performance on any dataset. We decide to not systematically tune all these parameters, as this is out of our scope, but to fix them for all the experiments, as done in the original network.\\

\textbf{Data pre-processing}. The signals are resampled to 128Hz and rescaled (per channel and per-subject), so that, for each channel, the EEG signal has median 0 and inter quartile range (IRQ) 1. The values with an absolute deviation from the median above 20*IQR are clipped. The signals outside the range of the scored hypnogram are trimmed. The recordings scored according to Rechtschaffen and Kales rules results in six scoring classes, \textit{i.e.}, awake, N1, N2, N3, N4, and REM. In order to use the AASM standard, we merge the N3 and N4 stages into a single stage N3. The loss function for stages as MOVEMENT and UNKNOWN is masked during the training procedure. \\

\textbf{Data sampling}. U-Sleep is trained using mini-batch Adam gradient-based optimizer. Each element in the batch is a sequence/segment of $L=35$ EEG and EOG 30-second signals/epochs from a single subject. Each sequence/element is sampled from the training data as follows. (1) dataset sampling: one dataset is selected randomly. The probability that a dataset $D$ is selected is given by {$P(D)=\alpha P\textsubscript{1}(D)+(1-\alpha)P\textsubscript{2}(D)$}, where $P\textsubscript{1}(D)$ is the probability that a dataset is sampled with a uniform distribution $1/N\textsubscript{D}$, where $N\textsubscript{D}$ is the number of available datasets, and $P\textsubscript{2}(D)$ is the probability of sampling a dataset according to its size. The parameter $\alpha$ is set to $0.5$ to equally weight $P\textsubscript{1}(D)$ and $P\textsubscript{2}(D)$; (2) subject sampling: a recording $S\textsubscript{D}$ is uniformly sampled from $D$; (3) channel sampling: one EEG and one EOG are uniformly sampled from the available combinations of channels in $S\textsubscript{D}$ (\textit{e.g.}, if 2 EEG and 2 EOG channels are available, four combinations are possible); {(4) segment sampling: a segment of EEG signal and a segment of EOG signal, both of length $L=35$, are selected as follows: first a class from ${W, N1, N2, N3, R}$ is uniformly sampled, then a 30-second epoch scored with the sampled class is selected randomly from the whole night recording, the chosen epoch is shifted into a random position of the segment of length $L$ and finally the sequence is extracted.}

\subsection*{Transfer learning}

We define transfer learning as in the following clear and simple statements: \\

"\textit{Transfer learning and domain adaptation refer to the situation where what has been learned in one setting (\textit{e.g.}, distribution $P_1$) is exploited to improve generalization in another setting (say, distribution $P_2$)}" \cite{goodfellow2016deep}; \\

"\textit{Given a source domain $D\textsubscript{S}$ and learning task $T\textsubscript{S}$, a target domain $D\textsubscript{T}$ and learning task $T\textsubscript{T}$, transfer learning aims to help improve the learning of the target predictive function $f\textsubscript{T}(\cdot)$ in $D\textsubscript{T}$ using the knowledge in $D\textsubscript{S}$ and $T\textsubscript{S}$, where $D\textsubscript{S}\neq D\textsubscript{T}$ and $T\textsubscript{S}\neq T\textsubscript{T}$}" \cite{pan2009survey}.\\

In our study the source and the target tasks are the same, i.e. $T\textsubscript{S}\equiv T\textsubscript{T}$. The task is always to perform sleep staging with the same set of sleep classes/stages. We want to transfer the knowledge about the previously learned sleep recordings (\textit{e.g.}, different hardware, different subject distributions with different sleep disorders) and the knowledge about the sleep scoring-rules (\textit{i.e.}, inter-scorer variability in the different data centers). The process generally involves overwriting a knowledge from a small-sized database to a previous big-sized knowledge (result of a long training process). One big concern is to avoid ending up in what the data scientists call catastrophic forgetting: "\textit{Also known as catastrophic interference, it is the tendency of an artificial neural network to completely and abruptly forget previously learned information upon learning new information}" as defined in \cite{mccloskey1989catastrophic}. Even if it is conceptually easy to understand, avoiding its occurrence is not trivial. To partially bypass this phenomena we fine-tune the architecture on the target domain using a smaller learning rate.\\

In our experiments we first pre-train the architecture on the data-source domain $S$ (\textit{e.g.}, a set of different domains/databases $\{S\textsubscript{DB\textsubscript{1}}, S\textsubscript{DB\textsubscript{2}}, . . ., S\textsubscript{DB\textsubscript{n}}\}$), then we fine-tune the model on the data-target domain $T$. Formally, we first minimize the loss function $L\textsubscript{S}$, resulting in the learned parameters $\theta$: 

\begin{equation}
argmin\textsubscript{$\theta$} = \sum_{(\textbf{x},\textbf{y})\in D\textsubscript{S}}^{} L\textsubscript{S}(\textbf{x}, P(\textbf{y}\|\textbf{x}), P\textsubscript{$\theta$}(\textbf{y}, \textbf{x}))
\label{eq_1}
\end{equation}

The parameters $\theta$ of the pre-trained model are used as the starting point on the data-target domain $T$. To transfer the learning on the new domain $T$, we fine-tune all the pre-trained parameters $\theta'=\theta$ (\textit{i.e.}, the entire network is further trained on the new data domain $T$):

\begin{equation}
argmin\textsubscript{$\theta'=\theta$} = \sum_{(\textbf{x},\textbf{y})\in D\textsubscript{T}}^{} L\textsubscript{T}(\textbf{x}, P(\textbf{y}\|\textbf{x}), P\textsubscript{$\theta$}(\textbf{y}, \textbf{x}))
\label{eq_2}
\end{equation}

\subsection*{Conditional learning}

Basically all the sleep scoring architectures learn in a conditional way. The aim is to maximize the conditional probability distributions $P(\textbf{Y}\vert\textbf{X})$, where $\textbf{X}$ are the sequences of the biosignals in input and $\textbf{Y}$ are the corresponding ground-truth labels. For each epoch $\textbf{x}_t$ in input the models aim to maximize the conditional probability distribution $P(y_t\vert\textbf{x}_t)$, where $y_t$ is the $t-th$ one-hot encoded vector of the ground-truth label. Hence, the model is trained to minimize the prediction error conditioned only by the knowledge of $\textbf{X}$. We know that the sleep data $\textbf{X}$ often come from different sources or data domains. Even in the same cohort, subjects with different demographics and sleep disorders may occur, resulting in significant shifts in their sleep data $\textbf{X}$ distributions. Imagine to have in the same data cohort $G$ different groups of subjects $\{g\textsubscript{1}, g\textsubscript{2}, . . ., g\textsubscript{G}\}$, with $g_\textsubscript{1}=\{healthy\}$, $g\textsubscript{2}=\{sleep\_apnea\}$ and so on. This additional information about the group (\textit{i.e.}, the sleep disorder group $g\textsubscript{i}$) to which the subject belongs can be given in input to the model.
So, we can either train $G$ fully separated models, each maximizing $G$ different $P(\textbf{Y}\vert\textbf{X})$ functions, or either train a single model maximizing the conditional probability distributions $P(\textbf{Y}\vert\textbf{X}, g\textsubscript{i})$. The latter - \textit{i.e.}, train the joint model with the additional condition $g\textsubscript{i}$ - is the smartest approach; the tasks are similar enough to benefit from sharing the parameters and the extracted features.

We decide to exploit the BN layers to insert the additional knowledge in the training of our model. In literature different normalization variants have been proposed by modulating the parameters of the vanilla BN layer \cite{dumoulin2016learned, de2017modulating, huang2017arbitrary, miyato2018spectral, xie2020adversarial}. We decide to exploit the sandwich batch normalization (SaBN) approach recently proposed in \cite{gong2022sandwich}.

The vanilla BN \cite{ioffe2015batch} normalizes the samples in a mini-batch in input by using the mean $\mu$ and the standard deviation $\sigma$, and then re-scales them with the $\gamma$ and $\beta$ parameters. So, given the feature in input $f \in \mathbb{R}^{B\times C\times H\times W}$, where $B$ is the batch size, $C$ is the number of channels and $H$ and $W$ are the height and width respectively, the vanilla BN computes: 

\begin{equation}
h = \gamma(\frac{f-\mu(f)}{\sigma(f)})+\beta
\label{eq_3}
\end{equation}

where $\mu(f)$ and $\sigma(f)$ are the mean and variance running estimates (batch statistics, \textit{i.e.}, moving mean and moving variance) computed on $f$ along $(N, H, W)$ dimensions; $\gamma$ and $\beta$ are the re-scaling learnable parameters of the BN affine layer with shape $C$. Clearly, the vanilla BN has only a single re-scaling transform, indirectly assuming all features coming from a single data distribution. In \cite{miyato2018spectral}, to tackle the data heterogeneity issue (\textit{i.e.}, images from different data domains/distributions), they propose the Categorical Conditional BN (CCBN), so boosting the quality of the generated images. The CCBN layer computes the following operation: 

\begin{equation}
h = \gamma\textsubscript{g}(\frac{f-\mu(f)}{\sigma(f)})+\beta\textsubscript{g} \qquad g=1, . . ., G
\label{eq_4}
\end{equation}

where $\gamma\textsubscript{g}$ and $\beta\textsubscript{g}$ are the re-scaling learnable parameters of each $g-th$ affine layer, where $g$ corresponds to the domain index associated to the input. The parameters of each affine layer are learned to capture the domain/distribution-specific information. In \cite{gong2022sandwich}, instead, they propose the SaBN layer, an improved variant of the CCBN. They claim that different individual affine layers might cause an imbalanced learning for the different domains/distributions. They factorize the BN affine layer into one shared "sandwich" BN layer cascaded by a set of independent BN affine layers, computed as follows:

\begin{equation}
h = \gamma\textsubscript{g}(\gamma\textsubscript{sa}(\frac{f-\mu(f)}{\sigma(f)})+\beta\textsubscript{sa})+\beta\textsubscript{g} \qquad i=1, . . ., G
\label{eq_5}
\end{equation}

where $\gamma\textsubscript{sa}$ and $\beta\textsubscript{sa}$ are the re-scaling learnable parameters of the "sandwich" shared affine BN layer, while, as above, $\gamma\textsubscript{g}$ and $\beta\textsubscript{g}$ are the re-scaling learnable parameters of each $g-th$ affine layer, conditioned on the categorical input $g$. The SaBN enable the conditional fine-tuning of a pre-trained U-Sleep architecture, conditioned by the categorical index in input $g$.

\subsection*{Evaluation}

In all our experiments we evaluate U-Sleep as stated in \cite{perslev2021u}. The model scores the full PSG, without considering the predicted class on a segment with a label different from the five sleep stages (\textit{e.g.}, segment labelled as 'UNKNOWN' or as 'MOVEMENT'). The final prediction is the results of all the possible combinations of the available EEG and EOG channels for each PSG. Hence, we use the majority vote, \textit{i.e.}, the ensemble of predictions given by the multiple combination of channels in input.\\

{The unweighted F1-score metric \cite{sokolova2009systematic} is computed on all the testing sets to evaluate the performance of the model on all the experiments. We compute the F1-score for all the five classes, we then combine them by calculating the unweighted mean. Note that the unweighted F1-scores reduce the absolute scores due to lower performance on less abundant classes such as sleep stage N1. For this reason, we also report in Supplementary Table 10, Supplementary Table 11, and  Supplementary Table 12 the results achieved in terms of weighted F1-score - \textit{i.e.}, the metric is weighted by the number of true instances for each label, so as to consider the high imbalance between the sleep stages. In that case, the absolute scores significantly increases on all the experiments. In Supplementary Table 10, Supplementary Table 11, and Supplementary Table 12 we also report the Cohen's kappa metric, given its valuable property of correcting the chance of agreement between the automatic sleep scoring algorithm, \textit{i.e.}, overall predicted sleep stages, and the ground truth, \textit{i.e.}, the sleep labels given by the physicians.}\\

* The Bern Sleep Data Base BSDB registry usage was ethically approved in the framework of the E12034 - SPAS (Sleep Physician Assistant System) Eurostar-Horizon 2020 program (Kantonale Ethikkommission Bern, 2020-01094).

\section*{Data availability}

{The Bern Sleep Data Base BSDB registry, the sleep disorder patient cohort of the Inselspital, University hospital Bern, is not publicly available. The BSDB data are available on request from the
corresponding author L.F. (legal conditions ensuring data privacy will be defined in a "data transfer agreement document", together with a description of the analysis project).} All other datasets are in principle publicly available, most datasets require the user to complete a data request form. The researchers and the use-case scenario need to be eligible for a given dataset. In Table~\ref{db_overview} we specify which datasets require approval from a Data Access Committee and which are directly available online.

\section*{Code availability}

The code we used in our study is based on what was previously developed in \cite{perslev2021u}, publicly available on GitHub at \url{https://github.com/perslev/U-Time}. All our experiments were carried out using the following branch \url{https://github.com/perslev/U-Time/tree/usleep-paper-version}. 
As a result of important feedback received from the whole community, but especially thanks to our important feedback related to the use of atypical and clinically non-recommended derivations, the authors provide the bugfixed code in \url{https://github.com/perslev/U-Time/tree/usleep-paper-version-branch-bugfixes}.

\section*{Acknowledgements}

F.D.F. was supported by SPAS: Sleep Physician Assistant System project, from Eurostars funding program. P.F. was supported by the Interfaculty Research Cooperation (IRC) Decoding Sleep: From Neurons to Health \& Mind, from the University of Bern, Switzerland. A.T. was supported by the IRC Decoding Sleep: From Neurons to Health \& Mind, from the University of Bern, and the Swiss National Science Foundation (\#320030\_188737).
The National Sleep Research Resource was supported by the National Heart, Lung, and Blood Institute (R24 HL114473, 75N92019R002). The Apnea, Bariatric surgery, and CPAP study (ABC Study) was supported by National Institutes of Health grants R01HL106410 and K24HL127307. Philips Respironics donated the CPAP machines and supplies used in the perioperative period for patients undergoing bariatric surgery. The Cleveland Children's Sleep and Health Study (CCSHS) was supported by grants from the National Institutes of Health (RO1HL60957, K23 HL04426, RO1 NR02707, M01 Rrmpd0380-39). The Cleveland Family Study (CFS) was supported by grants from the National Institutes of Health (HL46380, M01 RR00080-39, T32-HL07567, RO1-46380). The Childhood Adenotonsillectomy Trial (CHAT) was supported by the National Institutes of Health (HL083075, HL083129, UL1-RR-024134, UL1 RR024989). The Home Positive Airway Pressure study (HomePAP) was supported by the American Sleep Medicine Foundation 38-PM-07 Grant: Portable Monitoring for the Diagnosis and Management of OSA. The Multi-Ethnic Study of Atherosclerosis (MESA) Sleep Ancillary study was funded by NIH-NHLBI Association of Sleep Disorders with Cardiovascular Health Across Ethnic Groups (RO1 HL098433). MESA is supported by NHLBI funded contracts HHSN268201500003I, N01-HC-95159, N01-HC-95160, N01-HC-95161, N01-HC-95162, N01-HC-95163, N01-HC-95164, N01-HC-95165, N01-HC-95166, N01-HC-95167, N01-HC-95168 and N01-HC-95169 from the National Heart, Lung, and Blood Institute, and by cooperative agreements UL1-TR-000040, UL1-TR-001079, and UL1-TR-001420 funded by NCATS. The National Heart, Lung, and Blood Institute provided funding for the ancillary MrOS Sleep Study, "Outcomes of Sleep Disorders in Older Men," under the following grant numbers: R01 HL071194, R01 HL070848, R01 HL070847, R01 HL070842, R01 HL070841, R01 HL070837, R01 HL070838, and R01 HL070839. The Sleep Heart Health Study (SHHS) was supported by National Heart, Lung, and Blood Institute cooperative agreements U01HL53916 (University of California, Davis), U01HL53931 (New York University), U01HL53934 (University of Minnesota), U01HL53937 and U01HL64360 (Johns Hopkins University), U01HL53938 (University of Arizona), U01HL53940 (University of Washington), U01HL53941 (Boston University), and U01HL63463 (Case Western Reserve University).

\section*{Author contribution}

{All authors contributed to the design of the study; L.F. and G.M. contributed equally to the study, in particular they implemented the system and they conducted all the experiments; L.F., G.M. and F.D.F. wrote the paper with feedback from J.v.d.M, A.T., M.S. and P.F.; the BSDB dataset was extracted and prepared by J.v.d.M, M.P., L.F. and G.M.; all authors approved the final paper.}

\section*{Competing interests}

The authors declare no competing interests.

\section*{Additional information}

Publisher’s note Springer Nature remains neutral with regard to jurisdictional claims in published maps and institutional affiliations.




\begin{thebibliography}{76}
\ifx \bisbn   \undefined \def \bisbn  #1{ISBN #1}\fi
\ifx \binits  \undefined \def \binits#1{#1}\fi
\ifx \bauthor  \undefined \def \bauthor#1{#1}\fi
\ifx \batitle  \undefined \def \batitle#1{#1}\fi
\ifx \bjtitle  \undefined \def \bjtitle#1{#1}\fi
\ifx \bvolume  \undefined \def \bvolume#1{\textbf{#1}}\fi
\ifx \byear  \undefined \def \byear#1{#1}\fi
\ifx \bissue  \undefined \def \bissue#1{#1}\fi
\ifx \bfpage  \undefined \def \bfpage#1{#1}\fi
\ifx \blpage  \undefined \def \blpage #1{#1}\fi
\ifx \burl  \undefined \def \burl#1{\textsf{#1}}\fi
\ifx \doiurl  \undefined \def \doiurl#1{\url{https://doi.org/#1}}\fi
\ifx \betal  \undefined \def \betal{\textit{et al.}}\fi
\ifx \binstitute  \undefined \def \binstitute#1{#1}\fi
\ifx \binstitutionaled  \undefined \def \binstitutionaled#1{#1}\fi
\ifx \bctitle  \undefined \def \bctitle#1{#1}\fi
\ifx \beditor  \undefined \def \beditor#1{#1}\fi
\ifx \bpublisher  \undefined \def \bpublisher#1{#1}\fi
\ifx \bbtitle  \undefined \def \bbtitle#1{#1}\fi
\ifx \bedition  \undefined \def \bedition#1{#1}\fi
\ifx \bseriesno  \undefined \def \bseriesno#1{#1}\fi
\ifx \blocation  \undefined \def \blocation#1{#1}\fi
\ifx \bsertitle  \undefined \def \bsertitle#1{#1}\fi
\ifx \bsnm \undefined \def \bsnm#1{#1}\fi
\ifx \bsuffix \undefined \def \bsuffix#1{#1}\fi
\ifx \bparticle \undefined \def \bparticle#1{#1}\fi
\ifx \barticle \undefined \def \barticle#1{#1}\fi
\bibcommenthead
\ifx \bconfdate \undefined \def \bconfdate #1{#1}\fi
\ifx \botherref \undefined \def \botherref #1{#1}\fi
\ifx \url \undefined \def \url#1{\textsf{#1}}\fi
\ifx \bchapter \undefined \def \bchapter#1{#1}\fi
\ifx \bbook \undefined \def \bbook#1{#1}\fi
\ifx \bcomment \undefined \def \bcomment#1{#1}\fi
\ifx \oauthor \undefined \def \oauthor#1{#1}\fi
\ifx \citeauthoryear \undefined \def \citeauthoryear#1{#1}\fi
\ifx \endbibitem  \undefined \def \endbibitem {}\fi
\ifx \bconflocation  \undefined \def \bconflocation#1{#1}\fi
\ifx \arxivurl  \undefined \def \arxivurl#1{\textsf{#1}}\fi
\csname PreBibitemsHook\endcsname

\bibitem{berry2012rules}
\begin{barticle}
\bauthor{\bsnm{Berry}, \binits{R.B.}},
\bauthor{\bsnm{Budhiraja}, \binits{R.}},
\bauthor{\bsnm{Gottlieb}, \binits{D.J.}},
\bauthor{\bsnm{Gozal}, \binits{D.}},
\bauthor{\bsnm{Iber}, \binits{C.}},
\bauthor{\bsnm{Kapur}, \binits{V.K.}},
\bauthor{\bsnm{Marcus}, \binits{C.L.}},
\bauthor{\bsnm{Mehra}, \binits{R.}},
\bauthor{\bsnm{Parthasarathy}, \binits{S.}},
\bauthor{\bsnm{Quan}, \binits{S.F.}}, \betal:
\batitle{Rules for scoring respiratory events in sleep: update of the 2007 aasm
  manual for the scoring of sleep and associated events: deliberations of the
  sleep apnea definitions task force of the american academy of sleep
  medicine}.
\bjtitle{Journal of clinical sleep medicine}
\bvolume{8}(\bissue{5}),
\bfpage{597}--\blpage{619}
(\byear{2012})
\end{barticle}
\endbibitem

\bibitem{ronzhina2012sleep}
\begin{barticle}
\bauthor{\bsnm{Ronzhina}, \binits{M.}},
\bauthor{\bsnm{Janou{\v{s}}ek}, \binits{O.}},
\bauthor{\bsnm{Kol{\'a}{\v{r}}ov{\'a}}, \binits{J.}},
\bauthor{\bsnm{Nov{\'a}kov{\'a}}, \binits{M.}},
\bauthor{\bsnm{Honz{\'\i}k}, \binits{P.}},
\bauthor{\bsnm{Provazn{\'\i}k}, \binits{I.}}:
\batitle{Sleep scoring using artificial neural networks}.
\bjtitle{Sleep medicine reviews}
\bvolume{16}(\bissue{3}),
\bfpage{251}--\blpage{263}
(\byear{2012})
\end{barticle}
\endbibitem

\bibitem{csen2014comparative}
\begin{barticle}
\bauthor{\bsnm{{\c{S}}en}, \binits{B.}},
\bauthor{\bsnm{Peker}, \binits{M.}},
\bauthor{\bsnm{{\c{C}}avu{\c{s}}o{\u{g}}lu}, \binits{A.}},
\bauthor{\bsnm{{\c{C}}elebi}, \binits{F.V.}}:
\batitle{A comparative study on classification of sleep stage based on eeg
  signals using feature selection and classification algorithms}.
\bjtitle{Journal of medical systems}
\bvolume{38}(\bissue{3}),
\bfpage{18}
(\byear{2014})
\end{barticle}
\endbibitem

\bibitem{radha2014comparison}
\begin{bchapter}
\bauthor{\bsnm{Radha}, \binits{M.}},
\bauthor{\bsnm{Garcia-Molina}, \binits{G.}},
\bauthor{\bsnm{Poel}, \binits{M.}},
\bauthor{\bsnm{Tononi}, \binits{G.}}:
\bctitle{Comparison of feature and classifier algorithms for online automatic
  sleep staging based on a single eeg signal}.
In: \bbtitle{2014 36th Annual International Conference of the IEEE Engineering
  in Medicine and Biology Society},
pp. \bfpage{1876}--\blpage{1880}
(\byear{2014}).
\bcomment{IEEE}
\end{bchapter}
\endbibitem

\bibitem{aboalayon2016sleep}
\begin{barticle}
\bauthor{\bsnm{Aboalayon}, \binits{K.}},
\bauthor{\bsnm{Faezipour}, \binits{M.}},
\bauthor{\bsnm{Almuhammadi}, \binits{W.}},
\bauthor{\bsnm{Moslehpour}, \binits{S.}}:
\batitle{Sleep stage classification using eeg signal analysis: a comprehensive
  survey and new investigation}.
\bjtitle{Entropy}
\bvolume{18}(\bissue{9}),
\bfpage{272}
(\byear{2016})
\end{barticle}
\endbibitem

\bibitem{boostani2017comparative}
\begin{barticle}
\bauthor{\bsnm{Boostani}, \binits{R.}},
\bauthor{\bsnm{Karimzadeh}, \binits{F.}},
\bauthor{\bsnm{Nami}, \binits{M.}}:
\batitle{A comparative review on sleep stage classification methods in patients
  and healthy individuals}.
\bjtitle{Computer methods and programs in biomedicine}
\bvolume{140},
\bfpage{77}--\blpage{91}
(\byear{2017})
\end{barticle}
\endbibitem

\bibitem{fiorillo2019automated}
\begin{barticle}
\bauthor{\bsnm{Fiorillo}, \binits{L.}},
\bauthor{\bsnm{Puiatti}, \binits{A.}},
\bauthor{\bsnm{Papandrea}, \binits{M.}},
\bauthor{\bsnm{Ratti}, \binits{P.-L.}},
\bauthor{\bsnm{Favaro}, \binits{P.}},
\bauthor{\bsnm{Roth}, \binits{C.}},
\bauthor{\bsnm{Bargiotas}, \binits{P.}},
\bauthor{\bsnm{Bassetti}, \binits{C.L.}},
\bauthor{\bsnm{Faraci}, \binits{F.D.}}:
\batitle{Automated sleep scoring: a review of the latest approaches}.
\bjtitle{Sleep medicine reviews}
\bvolume{48},
\bfpage{101204}
(\byear{2019})
\end{barticle}
\endbibitem

\bibitem{tsinalis2016automatic}
\begin{barticle}
\bauthor{\bsnm{Tsinalis}, \binits{O.}},
\bauthor{\bsnm{Matthews}, \binits{P.M.}},
\bauthor{\bsnm{Guo}, \binits{Y.}}:
\batitle{Automatic sleep stage scoring using time-frequency analysis and
  stacked sparse autoencoders}.
\bjtitle{Annals of biomedical engineering}
\bvolume{44}(\bissue{5}),
\bfpage{1587}--\blpage{1597}
(\byear{2016})
\end{barticle}
\endbibitem

\bibitem{dong2018mixed}
\begin{barticle}
\bauthor{\bsnm{Dong}, \binits{H.}},
\bauthor{\bsnm{Supratak}, \binits{A.}},
\bauthor{\bsnm{Pan}, \binits{W.}},
\bauthor{\bsnm{Wu}, \binits{C.}},
\bauthor{\bsnm{Matthews}, \binits{P.M.}},
\bauthor{\bsnm{Guo}, \binits{Y.}}:
\batitle{Mixed neural network approach for temporal sleep stage
  classification}.
\bjtitle{IEEE Transactions on Neural Systems and Rehabilitation Engineering}
\bvolume{26}(\bissue{2}),
\bfpage{324}--\blpage{333}
(\byear{2018})
\end{barticle}
\endbibitem

\bibitem{perslev2019u}
\begin{botherref}
\oauthor{\bsnm{Perslev}, \binits{M.}},
\oauthor{\bsnm{Jensen}, \binits{M.}},
\oauthor{\bsnm{Darkner}, \binits{S.}},
\oauthor{\bsnm{Jennum}, \binits{P.J.}},
\oauthor{\bsnm{Igel}, \binits{C.}}:
U-time: A fully convolutional network for time series segmentation applied to
  sleep staging.
Advances in Neural Information Processing Systems
\textbf{32}
(2019)
\end{botherref}
\endbibitem

\bibitem{perslev2021u}
\begin{barticle}
\bauthor{\bsnm{Perslev}, \binits{M.}},
\bauthor{\bsnm{Darkner}, \binits{S.}},
\bauthor{\bsnm{Kempfner}, \binits{L.}},
\bauthor{\bsnm{Nikolic}, \binits{M.}},
\bauthor{\bsnm{Jennum}, \binits{P.J.}},
\bauthor{\bsnm{Igel}, \binits{C.}}:
\batitle{U-sleep: resilient high-frequency sleep staging}.
\bjtitle{NPJ digital medicine}
\bvolume{4}(\bissue{1}),
\bfpage{1}--\blpage{12}
(\byear{2021})
\end{barticle}
\endbibitem

\bibitem{tsinalis2016automaticCNN}
\begin{botherref}
\oauthor{\bsnm{Tsinalis}, \binits{O.}},
\oauthor{\bsnm{Matthews}, \binits{P.M.}},
\oauthor{\bsnm{Guo}, \binits{Y.}},
\oauthor{\bsnm{Zafeiriou}, \binits{S.}}:
Automatic sleep stage scoring with single-channel EEG using convolutional
  neural networks.
Preprint at \url{https://arxiv.org/abs/1610.01683}
(2016)
\end{botherref}
\endbibitem

\bibitem{vilamala2017deep}
\begin{bchapter}
\bauthor{\bsnm{Vilamala}, \binits{A.}},
\bauthor{\bsnm{Madsen}, \binits{K.H.}},
\bauthor{\bsnm{Hansen}, \binits{L.K.}}:
\bctitle{Deep convolutional neural networks for interpretable analysis of eeg
  sleep stage scoring}.
In: \bbtitle{2017 IEEE 27th International Workshop on Machine Learning for
  Signal Processing (MLSP)},
pp. \bfpage{1}--\blpage{6}
(\byear{2017}).
\bcomment{IEEE}
\end{bchapter}
\endbibitem

\bibitem{zhang2018complex}
\begin{barticle}
\bauthor{\bsnm{Zhang}, \binits{J.}},
\bauthor{\bsnm{Wu}, \binits{Y.}}:
\batitle{Complex-valued unsupervised convolutional neural networks for sleep
  stage classification}.
\bjtitle{Computer methods and programs in biomedicine}
\bvolume{164},
\bfpage{181}--\blpage{191}
(\byear{2018})
\end{barticle}
\endbibitem

\bibitem{chambon2018deep}
\begin{barticle}
\bauthor{\bsnm{Chambon}, \binits{S.}},
\bauthor{\bsnm{Galtier}, \binits{M.N.}},
\bauthor{\bsnm{Arnal}, \binits{P.J.}},
\bauthor{\bsnm{Wainrib}, \binits{G.}},
\bauthor{\bsnm{Gramfort}, \binits{A.}}:
\batitle{A deep learning architecture for temporal sleep stage classification
  using multivariate and multimodal time series}.
\bjtitle{IEEE Transactions on Neural Systems and Rehabilitation Engineering}
\bvolume{26}(\bissue{4}),
\bfpage{758}--\blpage{769}
(\byear{2018})
\end{barticle}
\endbibitem

\bibitem{cui2018automatic}
\begin{botherref}
\oauthor{\bsnm{Cui}, \binits{Z.}},
\oauthor{\bsnm{Zheng}, \binits{X.}},
\oauthor{\bsnm{Shao}, \binits{X.}},
\oauthor{\bsnm{Cui}, \binits{L.}}:
Automatic sleep stage classification based on convolutional neural network and
  fine-grained segments.
Complexity
\textbf{2018}
(2018)
\end{botherref}
\endbibitem

\bibitem{olesen2018deep}
\begin{bchapter}
\bauthor{\bsnm{Olesen}, \binits{A.N.}},
\bauthor{\bsnm{Jennum}, \binits{P.}},
\bauthor{\bsnm{Peppard}, \binits{P.}},
\bauthor{\bsnm{Mignot}, \binits{E.}},
\bauthor{\bsnm{Sorensen}, \binits{H.B.}}:
\bctitle{Deep residual networks for automatic sleep stage classification of raw
  polysomnographic waveforms}.
In: \bbtitle{2018 40th Annual International Conference of the IEEE Engineering
  in Medicine and Biology Society (EMBC)},
pp. \bfpage{1}--\blpage{4}
(\byear{2018}).
\bcomment{IEEE}
\end{bchapter}
\endbibitem

\bibitem{patanaik2018end}
\begin{barticle}
\bauthor{\bsnm{Patanaik}, \binits{A.}},
\bauthor{\bsnm{Ong}, \binits{J.L.}},
\bauthor{\bsnm{Gooley}, \binits{J.J.}},
\bauthor{\bsnm{Ancoli-Israel}, \binits{S.}},
\bauthor{\bsnm{Chee}, \binits{M.W.}}:
\batitle{An end-to-end framework for real-time automatic sleep stage
  classification}.
\bjtitle{Sleep}
\bvolume{41}(\bissue{5}),
\bfpage{041}
(\byear{2018})
\end{barticle}
\endbibitem

\bibitem{sors2018convolutional}
\begin{barticle}
\bauthor{\bsnm{Sors}, \binits{A.}},
\bauthor{\bsnm{Bonnet}, \binits{S.}},
\bauthor{\bsnm{Mirek}, \binits{S.}},
\bauthor{\bsnm{Vercueil}, \binits{L.}},
\bauthor{\bsnm{Payen}, \binits{J.-F.}}:
\batitle{A convolutional neural network for sleep stage scoring from raw
  single-channel eeg}.
\bjtitle{Biomedical Signal Processing and Control}
\bvolume{42},
\bfpage{107}--\blpage{114}
(\byear{2018})
\end{barticle}
\endbibitem

\bibitem{yildirim2019deep}
\begin{barticle}
\bauthor{\bsnm{Yildirim}, \binits{O.}},
\bauthor{\bsnm{Baloglu}, \binits{U.B.}},
\bauthor{\bsnm{Acharya}, \binits{U.R.}}:
\batitle{A deep learning model for automated sleep stages classification using
  psg signals}.
\bjtitle{International journal of environmental research and public health}
\bvolume{16}(\bissue{4}),
\bfpage{599}
(\byear{2019})
\end{barticle}
\endbibitem

\bibitem{fiorillo2020temporal}
\begin{bchapter}
\bauthor{\bsnm{Fiorillo}, \binits{L.}},
\bauthor{\bsnm{Wand}, \binits{M.}},
\bauthor{\bsnm{Marino}, \binits{I.}},
\bauthor{\bsnm{Favaro}, \binits{P.}},
\bauthor{\bsnm{Faraci}, \binits{F.D.}}:
\bctitle{Temporal dependency in automatic sleep scoring via deep learning based
  architectures: An empirical study}.
In: \bbtitle{2020 42nd Annual International Conference of the IEEE Engineering
  in Medicine \& Biology Society (EMBC)},
pp. \bfpage{3509}--\blpage{3512}
(\byear{2020}).
\bcomment{IEEE}
\end{bchapter}
\endbibitem

\bibitem{michielli2019cascaded}
\begin{barticle}
\bauthor{\bsnm{Michielli}, \binits{N.}},
\bauthor{\bsnm{Acharya}, \binits{U.R.}},
\bauthor{\bsnm{Molinari}, \binits{F.}}:
\batitle{Cascaded lstm recurrent neural network for automated sleep stage
  classification using single-channel eeg signals}.
\bjtitle{Computers in biology and medicine}
\bvolume{106},
\bfpage{71}--\blpage{81}
(\byear{2019})
\end{barticle}
\endbibitem

\bibitem{phan2019seqsleepnet}
\begin{botherref}
\oauthor{\bsnm{{Phan}}, \binits{H.}},
\oauthor{\bsnm{{Andreotti}}, \binits{F.}},
\oauthor{\bsnm{{Cooray}}, \binits{N.}},
\oauthor{\bsnm{{Chén}}, \binits{O.Y.}},
\oauthor{\bsnm{{De Vos}}, \binits{M.}}:
Seqsleepnet: End-to-end hierarchical recurrent neural network for
  sequence-to-sequence automatic sleep staging.
IEEE Transactions on Neural Systems and Rehabilitation Engineering,
1
(2019)
\end{botherref}
\endbibitem

\bibitem{supratak2017deepsleepnet}
\begin{barticle}
\bauthor{\bsnm{Supratak}, \binits{A.}},
\bauthor{\bsnm{Dong}, \binits{H.}},
\bauthor{\bsnm{Wu}, \binits{C.}},
\bauthor{\bsnm{Guo}, \binits{Y.}}:
\batitle{Deepsleepnet: a model for automatic sleep stage scoring based on raw
  single-channel eeg}.
\bjtitle{IEEE Transactions on Neural Systems and Rehabilitation Engineering}
\bvolume{25}(\bissue{11}),
\bfpage{1998}--\blpage{2008}
(\byear{2017})
\end{barticle}
\endbibitem

\bibitem{biswal2018expert}
\begin{barticle}
\bauthor{\bsnm{Biswal}, \binits{S.}},
\bauthor{\bsnm{Sun}, \binits{H.}},
\bauthor{\bsnm{Goparaju}, \binits{B.}},
\bauthor{\bsnm{Westover}, \binits{M.B.}},
\bauthor{\bsnm{Sun}, \binits{J.}},
\bauthor{\bsnm{Bianchi}, \binits{M.T.}}:
\batitle{Expert-level sleep scoring with deep neural networks}.
\bjtitle{Journal of the American Medical Informatics Association}
\bvolume{25}(\bissue{12}),
\bfpage{1643}--\blpage{1650}
(\byear{2018})
\end{barticle}
\endbibitem

\bibitem{malafeev2018automatic}
\begin{barticle}
\bauthor{\bsnm{Malafeev}, \binits{A.}},
\bauthor{\bsnm{Laptev}, \binits{D.}},
\bauthor{\bsnm{Bauer}, \binits{S.}},
\bauthor{\bsnm{Omlin}, \binits{X.}},
\bauthor{\bsnm{Wierzbicka}, \binits{A.}},
\bauthor{\bsnm{Wichniak}, \binits{A.}},
\bauthor{\bsnm{Jernajczyk}, \binits{W.}},
\bauthor{\bsnm{Riener}, \binits{R.}},
\bauthor{\bsnm{Buhmann}, \binits{J.M.}},
\bauthor{\bsnm{Achermann}, \binits{P.}}:
\batitle{Automatic human sleep stage scoring using deep neural networks}.
\bjtitle{Frontiers in Neuroscience}
\bvolume{12},
\bfpage{781}
(\byear{2018})
\end{barticle}
\endbibitem

\bibitem{stephansen2018neural}
\begin{barticle}
\bauthor{\bsnm{Stephansen}, \binits{J.B.}},
\bauthor{\bsnm{Olesen}, \binits{A.N.}},
\bauthor{\bsnm{Olsen}, \binits{M.}},
\bauthor{\bsnm{Ambati}, \binits{A.}},
\bauthor{\bsnm{Leary}, \binits{E.B.}},
\bauthor{\bsnm{Moore}, \binits{H.E.}},
\bauthor{\bsnm{Carrillo}, \binits{O.}},
\bauthor{\bsnm{Lin}, \binits{L.}},
\bauthor{\bsnm{Han}, \binits{F.}},
\bauthor{\bsnm{Yan}, \binits{H.}}, \betal:
\batitle{Neural network analysis of sleep stages enables efficient diagnosis of
  narcolepsy}.
\bjtitle{Nature communications}
\bvolume{9}(\bissue{1}),
\bfpage{5229}
(\byear{2018})
\end{barticle}
\endbibitem

\bibitem{back2019intra}
\begin{botherref}
\oauthor{\bsnm{Back}, \binits{S.}},
\oauthor{\bsnm{Lee}, \binits{S.}},
\oauthor{\bsnm{Seo}, \binits{H.}},
\oauthor{\bsnm{Park}, \binits{D.}},
\oauthor{\bsnm{Kim}, \binits{T.}},
\oauthor{\bsnm{Lee}, \binits{K.}}:
Intra-and inter-epoch temporal context network (IITNet) for automatic sleep
  stage scoring.
Preprint at \url{https://arxiv.org/abs/1902.06562}
(2019)
\end{botherref}
\endbibitem

\bibitem{mousavi2019sleepeegnet}
\begin{barticle}
\bauthor{\bsnm{Mousavi}, \binits{S.}},
\bauthor{\bsnm{Afghah}, \binits{F.}},
\bauthor{\bsnm{Acharya}, \binits{U.R.}}:
\batitle{Sleepeegnet: Automated sleep stage scoring with sequence to sequence
  deep learning approach}.
\bjtitle{PloS one}
\bvolume{14}(\bissue{5}),
\bfpage{0216456}
(\byear{2019})
\end{barticle}
\endbibitem

\bibitem{seo2020intra}
\begin{barticle}
\bauthor{\bsnm{Seo}, \binits{H.}},
\bauthor{\bsnm{Back}, \binits{S.}},
\bauthor{\bsnm{Lee}, \binits{S.}},
\bauthor{\bsnm{Park}, \binits{D.}},
\bauthor{\bsnm{Kim}, \binits{T.}},
\bauthor{\bsnm{Lee}, \binits{K.}}:
\batitle{Intra-and inter-epoch temporal context network (iitnet) using
  sub-epoch features for automatic sleep scoring on raw single-channel eeg}.
\bjtitle{Biomedical Signal Processing and Control}
\bvolume{61},
\bfpage{102037}
(\byear{2020})
\end{barticle}
\endbibitem

\bibitem{phan2020towards}
\begin{barticle}
\bauthor{\bsnm{Phan}, \binits{H.}},
\bauthor{\bsnm{Ch{\'e}n}, \binits{O.Y.}},
\bauthor{\bsnm{Koch}, \binits{P.}},
\bauthor{\bsnm{Lu}, \binits{Z.}},
\bauthor{\bsnm{McLoughlin}, \binits{I.}},
\bauthor{\bsnm{Mertins}, \binits{A.}},
\bauthor{\bsnm{De~Vos}, \binits{M.}}:
\batitle{Towards more accurate automatic sleep staging via deep transfer
  learning}.
\bjtitle{IEEE Transactions on Biomedical Engineering}
\bvolume{68}(\bissue{6}),
\bfpage{1787}--\blpage{1798}
(\byear{2020})
\end{barticle}
\endbibitem

\bibitem{supratak2020tinysleepnet}
\begin{bchapter}
\bauthor{\bsnm{Supratak}, \binits{A.}},
\bauthor{\bsnm{Guo}, \binits{Y.}}:
\bctitle{Tinysleepnet: An efficient deep learning model for sleep stage scoring
  based on raw single-channel eeg}.
In: \bbtitle{2020 42nd Annual International Conference of the IEEE Engineering
  in Medicine \& Biology Society (EMBC)},
pp. \bfpage{641}--\blpage{644}
(\byear{2020}).
\bcomment{IEEE}
\end{bchapter}
\endbibitem

\bibitem{phan2021xsleepnet}
\begin{botherref}
\oauthor{\bsnm{Phan}, \binits{H.}},
\oauthor{\bsnm{Ch{\'e}n}, \binits{O.Y.}},
\oauthor{\bsnm{Tran}, \binits{M.C.}},
\oauthor{\bsnm{Koch}, \binits{P.}},
\oauthor{\bsnm{Mertins}, \binits{A.}},
\oauthor{\bsnm{De~Vos}, \binits{M.}}:
Xsleepnet: Multi-view sequential model for automatic sleep staging.
IEEE Transactions on Pattern Analysis and Machine Intelligence
(2021)
\end{botherref}
\endbibitem

\bibitem{rechtschaffen1968manual}
\begin{botherref}
\oauthor{\bsnm{Rechtschaffen}, \binits{A.}},
\oauthor{\bsnm{Kales}, \binits{A.}}:
A Manual of Standardized Terminology, Techniques and Scoring System for Sleep
  Stages of Human Subjects.
(1968)
\end{botherref}
\endbibitem

\bibitem{huber2004local}
\begin{barticle}
\bauthor{\bsnm{Huber}, \binits{R.}},
\bauthor{\bsnm{Felice~Ghilardi}, \binits{M.}},
\bauthor{\bsnm{Massimini}, \binits{M.}},
\bauthor{\bsnm{Tononi}, \binits{G.}}:
\batitle{Local sleep and learning}.
\bjtitle{Nature}
\bvolume{430}(\bissue{6995}),
\bfpage{78}--\blpage{81}
(\byear{2004})
\end{barticle}
\endbibitem

\bibitem{nakamura2019hearables}
\begin{barticle}
\bauthor{\bsnm{Nakamura}, \binits{T.}},
\bauthor{\bsnm{Alqurashi}, \binits{Y.D.}},
\bauthor{\bsnm{Morrell}, \binits{M.J.}},
\bauthor{\bsnm{Mandic}, \binits{D.P.}}:
\batitle{Hearables: automatic overnight sleep monitoring with standardized
  in-ear eeg sensor}.
\bjtitle{IEEE Transactions on Biomedical Engineering}
\bvolume{67}(\bissue{1}),
\bfpage{203}--\blpage{212}
(\byear{2019})
\end{barticle}
\endbibitem

\bibitem{mikkelsen2021sleep}
\begin{barticle}
\bauthor{\bsnm{Mikkelsen}, \binits{K.B.}},
\bauthor{\bsnm{Phan}, \binits{H.}},
\bauthor{\bsnm{Rank}, \binits{M.L.}},
\bauthor{\bsnm{Hemmsen}, \binits{M.C.}},
\bauthor{\bsnm{De~Vos}, \binits{M.}},
\bauthor{\bsnm{Kidmose}, \binits{P.}}:
\batitle{Sleep monitoring using ear-centered setups: Investigating the
  influence from electrode configurations}.
\bjtitle{IEEE Transactions on Biomedical Engineering}
\bvolume{69}(\bissue{5}),
\bfpage{1564}--\blpage{1572}
(\byear{2021})
\end{barticle}
\endbibitem

\bibitem{jorgensen2020ear}
\begin{barticle}
\bauthor{\bsnm{J{\o}rgensen}, \binits{S.D.}},
\bauthor{\bsnm{Zibrandtsen}, \binits{I.C.}},
\bauthor{\bsnm{Kjaer}, \binits{T.W.}}:
\batitle{Ear-eeg-based sleep scoring in epilepsy: A comparison with scalp-eeg}.
\bjtitle{Journal of Sleep Research}
\bvolume{29}(\bissue{6}),
\bfpage{12921}
(\byear{2020})
\end{barticle}
\endbibitem

\bibitem{Ohayon2004}
\begin{barticle}
\bauthor{\bsnm{Ohayon}, \binits{M.}},
\bauthor{\bsnm{Carskadon}, \binits{M.}},
\bauthor{\bsnm{Guilleminault}, \binits{C.}},
\bauthor{\bsnm{Vitiello}, \binits{M.}}:
\batitle{Meta-analysis of quantitative sleep parameters from childhood to old
  age in healthy individuals: Developing normative sleep values across the
  human lifespan}.
\bjtitle{Sleep}
\bvolume{27},
\bfpage{1255}--\blpage{73}
(\byear{2004})
\end{barticle}
\endbibitem

\bibitem{kocevska2021sleep}
\begin{barticle}
\bauthor{\bsnm{Kocevska}, \binits{D.}},
\bauthor{\bsnm{Lysen}, \binits{T.S.}},
\bauthor{\bsnm{Dotinga}, \binits{A.}},
\bauthor{\bsnm{Koopman-Verhoeff}, \binits{M.E.}},
\bauthor{\bsnm{Luijk}, \binits{M.P.}},
\bauthor{\bsnm{Antypa}, \binits{N.}},
\bauthor{\bsnm{Biermasz}, \binits{N.R.}},
\bauthor{\bsnm{Blokstra}, \binits{A.}},
\bauthor{\bsnm{Brug}, \binits{J.}},
\bauthor{\bsnm{Burk}, \binits{W.J.}}, \betal:
\batitle{Sleep characteristics across the lifespan in 1.1 million people from
  the netherlands, united kingdom and united states: a systematic review and
  meta-analysis}.
\bjtitle{Nature human behaviour}
\bvolume{5}(\bissue{1}),
\bfpage{113}--\blpage{122}
(\byear{2021})
\end{barticle}
\endbibitem

\bibitem{guillot2021robustsleepnet}
\begin{barticle}
\bauthor{\bsnm{Guillot}, \binits{A.}},
\bauthor{\bsnm{Thorey}, \binits{V.}}:
\batitle{Robustsleepnet: Transfer learning for automated sleep staging at
  scale}.
\bjtitle{IEEE Transactions on Neural Systems and Rehabilitation Engineering}
\bvolume{29},
\bfpage{1441}--\blpage{1451}
(\byear{2021})
\end{barticle}
\endbibitem

\bibitem{olesen2021automatic}
\begin{barticle}
\bauthor{\bsnm{Olesen}, \binits{A.N.}},
\bauthor{\bsnm{J{\o}rgen~Jennum}, \binits{P.}},
\bauthor{\bsnm{Mignot}, \binits{E.}},
\bauthor{\bsnm{Sorensen}, \binits{H.B.D.}}:
\batitle{Automatic sleep stage classification with deep residual networks in a
  mixed-cohort setting}.
\bjtitle{Sleep}
\bvolume{44}(\bissue{1}),
\bfpage{161}
(\byear{2021})
\end{barticle}
\endbibitem

\bibitem{vallat2021open}
\begin{barticle}
\bauthor{\bsnm{Vallat}, \binits{R.}},
\bauthor{\bsnm{Walker}, \binits{M.P.}}:
\batitle{An open-source, high-performance tool for automated sleep staging}.
\bjtitle{Elife}
\bvolume{10},
\bfpage{70092}
(\byear{2021})
\end{barticle}
\endbibitem

\bibitem{mathis2022u}
\begin{botherref}
\oauthor{\bsnm{Mathis}, \binits{J.}},
\oauthor{\bsnm{Andres}, \binits{D.}},
\oauthor{\bsnm{Schmitt}, \binits{W.}},
\oauthor{\bsnm{Bassetti}, \binits{C.}},
\oauthor{\bsnm{Hess}, \binits{C.}},
\oauthor{\bsnm{Schreier}, \binits{D.}}:
The diagnostic value of sleep and vigilance tests in central disorders of
  hypersomnolence.
Sleep
\textbf{45(3)}
(2022)
\end{botherref}
\endbibitem

\bibitem{zhang2018national}
\begin{barticle}
\bauthor{\bsnm{Zhang}, \binits{G.-Q.}},
\bauthor{\bsnm{Cui}, \binits{L.}},
\bauthor{\bsnm{Mueller}, \binits{R.}},
\bauthor{\bsnm{Tao}, \binits{S.}},
\bauthor{\bsnm{Kim}, \binits{M.}},
\bauthor{\bsnm{Rueschman}, \binits{M.}},
\bauthor{\bsnm{Mariani}, \binits{S.}},
\bauthor{\bsnm{Mobley}, \binits{D.}},
\bauthor{\bsnm{Redline}, \binits{S.}}:
\batitle{The national sleep research resource: towards a sleep data commons}.
\bjtitle{Journal of the American Medical Informatics Association}
\bvolume{25}(\bissue{10}),
\bfpage{1351}--\blpage{1358}
(\byear{2018})
\end{barticle}
\endbibitem

\bibitem{bakker2018gastric}
\begin{barticle}
\bauthor{\bsnm{Bakker}, \binits{J.P.}},
\bauthor{\bsnm{Tavakkoli}, \binits{A.}},
\bauthor{\bsnm{Rueschman}, \binits{M.}},
\bauthor{\bsnm{Wang}, \binits{W.}},
\bauthor{\bsnm{Andrews}, \binits{R.}},
\bauthor{\bsnm{Malhotra}, \binits{A.}},
\bauthor{\bsnm{Owens}, \binits{R.L.}},
\bauthor{\bsnm{Anand}, \binits{A.}},
\bauthor{\bsnm{Dudley}, \binits{K.A.}},
\bauthor{\bsnm{Patel}, \binits{S.R.}}:
\batitle{Gastric banding surgery versus continuous positive airway pressure for
  obstructive sleep apnea: a randomized controlled trial}.
\bjtitle{American journal of respiratory and critical care medicine}
\bvolume{197}(\bissue{8}),
\bfpage{1080}--\blpage{1083}
(\byear{2018})
\end{barticle}
\endbibitem

\bibitem{rosen2003prevalence}
\begin{barticle}
\bauthor{\bsnm{Rosen}, \binits{C.L.}},
\bauthor{\bsnm{Larkin}, \binits{E.K.}},
\bauthor{\bsnm{Kirchner}, \binits{H.L.}},
\bauthor{\bsnm{Emancipator}, \binits{J.L.}},
\bauthor{\bsnm{Bivins}, \binits{S.F.}},
\bauthor{\bsnm{Surovec}, \binits{S.A.}},
\bauthor{\bsnm{Martin}, \binits{R.J.}},
\bauthor{\bsnm{Redline}, \binits{S.}}:
\batitle{Prevalence and risk factors for sleep-disordered breathing in 8-to
  11-year-old children: association with race and prematurity}.
\bjtitle{The Journal of pediatrics}
\bvolume{142}(\bissue{4}),
\bfpage{383}--\blpage{389}
(\byear{2003})
\end{barticle}
\endbibitem

\bibitem{redline1995familial}
\begin{barticle}
\bauthor{\bsnm{Redline}, \binits{S.}},
\bauthor{\bsnm{Tishler}, \binits{P.V.}},
\bauthor{\bsnm{Tosteson}, \binits{T.D.}},
\bauthor{\bsnm{Williamson}, \binits{J.}},
\bauthor{\bsnm{Kump}, \binits{K.}},
\bauthor{\bsnm{Browner}, \binits{I.}},
\bauthor{\bsnm{Ferrette}, \binits{V.}},
\bauthor{\bsnm{Krejci}, \binits{P.}}:
\batitle{The familial aggregation of obstructive sleep apnea.}
\bjtitle{American journal of respiratory and critical care medicine}
\bvolume{151}(\bissue{3}),
\bfpage{682}--\blpage{687}
(\byear{1995})
\end{barticle}
\endbibitem

\bibitem{marcus2013randomized}
\begin{barticle}
\bauthor{\bsnm{Marcus}, \binits{C.L.}},
\bauthor{\bsnm{Moore}, \binits{R.H.}},
\bauthor{\bsnm{Rosen}, \binits{C.L.}},
\bauthor{\bsnm{Giordani}, \binits{B.}},
\bauthor{\bsnm{Garetz}, \binits{S.L.}},
\bauthor{\bsnm{Taylor}, \binits{H.G.}},
\bauthor{\bsnm{Mitchell}, \binits{R.B.}},
\bauthor{\bsnm{Amin}, \binits{R.}},
\bauthor{\bsnm{Katz}, \binits{E.S.}},
\bauthor{\bsnm{Arens}, \binits{R.}}, \betal:
\batitle{A randomized trial of adenotonsillectomy for childhood sleep apnea}.
\bjtitle{N Engl J Med}
\bvolume{368},
\bfpage{2366}--\blpage{2376}
(\byear{2013})
\end{barticle}
\endbibitem

\bibitem{redline2011childhood}
\begin{barticle}
\bauthor{\bsnm{Redline}, \binits{S.}},
\bauthor{\bsnm{Amin}, \binits{R.}},
\bauthor{\bsnm{Beebe}, \binits{D.}},
\bauthor{\bsnm{Chervin}, \binits{R.D.}},
\bauthor{\bsnm{Garetz}, \binits{S.L.}},
\bauthor{\bsnm{Giordani}, \binits{B.}},
\bauthor{\bsnm{Marcus}, \binits{C.L.}},
\bauthor{\bsnm{Moore}, \binits{R.H.}},
\bauthor{\bsnm{Rosen}, \binits{C.L.}},
\bauthor{\bsnm{Arens}, \binits{R.}}, \betal:
\batitle{The childhood adenotonsillectomy trial (chat): rationale, design, and
  challenges of a randomized controlled trial evaluating a standard surgical
  procedure in a pediatric population}.
\bjtitle{Sleep}
\bvolume{34}(\bissue{11}),
\bfpage{1509}--\blpage{1517}
(\byear{2011})
\end{barticle}
\endbibitem

\bibitem{rosen2012multisite}
\begin{barticle}
\bauthor{\bsnm{Rosen}, \binits{C.L.}},
\bauthor{\bsnm{Auckley}, \binits{D.}},
\bauthor{\bsnm{Benca}, \binits{R.}},
\bauthor{\bsnm{Foldvary-Schaefer}, \binits{N.}},
\bauthor{\bsnm{Iber}, \binits{C.}},
\bauthor{\bsnm{Kapur}, \binits{V.}},
\bauthor{\bsnm{Rueschman}, \binits{M.}},
\bauthor{\bsnm{Zee}, \binits{P.}},
\bauthor{\bsnm{Redline}, \binits{S.}}:
\batitle{A multisite randomized trial of portable sleep studies and positive
  airway pressure autotitration versus laboratory-based polysomnography for the
  diagnosis and treatment of obstructive sleep apnea: the homepap study}.
\bjtitle{Sleep}
\bvolume{35}(\bissue{6}),
\bfpage{757}--\blpage{767}
(\byear{2012})
\end{barticle}
\endbibitem

\bibitem{chen2015racial}
\begin{barticle}
\bauthor{\bsnm{Chen}, \binits{X.}},
\bauthor{\bsnm{Wang}, \binits{R.}},
\bauthor{\bsnm{Zee}, \binits{P.}},
\bauthor{\bsnm{Lutsey}, \binits{P.L.}},
\bauthor{\bsnm{Javaheri}, \binits{S.}},
\bauthor{\bsnm{Alc{\'a}ntara}, \binits{C.}},
\bauthor{\bsnm{Jackson}, \binits{C.L.}},
\bauthor{\bsnm{Williams}, \binits{M.A.}},
\bauthor{\bsnm{Redline}, \binits{S.}}:
\batitle{Racial/ethnic differences in sleep disturbances: the multi-ethnic
  study of atherosclerosis (mesa)}.
\bjtitle{Sleep}
\bvolume{38}(\bissue{6}),
\bfpage{877}--\blpage{888}
(\byear{2015})
\end{barticle}
\endbibitem

\bibitem{blackwell2011associations}
\begin{barticle}
\bauthor{\bsnm{Blackwell}, \binits{T.}},
\bauthor{\bsnm{Yaffe}, \binits{K.}},
\bauthor{\bsnm{Ancoli-Israel}, \binits{S.}},
\bauthor{\bsnm{Redline}, \binits{S.}},
\bauthor{\bsnm{Ensrud}, \binits{K.E.}},
\bauthor{\bsnm{Stefanick}, \binits{M.L.}},
\bauthor{\bsnm{Laffan}, \binits{A.}},
\bauthor{\bsnm{Stone}, \binits{K.L.}},
\bauthor{\bparticle{in} \bsnm{Men Study~Group}, \binits{O.F.}}:
\batitle{Associations between sleep architecture and sleep-disordered breathing
  and cognition in older community-dwelling men: the osteoporotic fractures in
  men sleep study}.
\bjtitle{Journal of the American Geriatrics Society}
\bvolume{59}(\bissue{12}),
\bfpage{2217}--\blpage{2225}
(\byear{2011})
\end{barticle}
\endbibitem

\bibitem{osteoporotic2015relationships}
\begin{botherref}
Relationships between sleep stages and changes in cognitive function in older
  men: the mros sleep study.
Sleep
\textbf{38}(3),
411--421
(2015)
\end{botherref}
\endbibitem

\bibitem{goldberger2000physiobank}
\begin{barticle}
\bauthor{\bsnm{Goldberger}, \binits{A.L.}},
\bauthor{\bsnm{Amaral}, \binits{L.A.}},
\bauthor{\bsnm{Glass}, \binits{L.}},
\bauthor{\bsnm{Hausdorff}, \binits{J.M.}},
\bauthor{\bsnm{Ivanov}, \binits{P.C.}},
\bauthor{\bsnm{Mark}, \binits{R.G.}},
\bauthor{\bsnm{Mietus}, \binits{J.E.}},
\bauthor{\bsnm{Moody}, \binits{G.B.}},
\bauthor{\bsnm{Peng}, \binits{C.-K.}},
\bauthor{\bsnm{Stanley}, \binits{H.E.}}:
\batitle{Physiobank, physiotoolkit, and physionet: components of a new research
  resource for complex physiologic signals}.
\bjtitle{circulation}
\bvolume{101}(\bissue{23}),
\bfpage{215}--\blpage{220}
(\byear{2000})
\end{barticle}
\endbibitem

\bibitem{ghassemi2018you}
\begin{bchapter}
\bauthor{\bsnm{Ghassemi}, \binits{M.M.}},
\bauthor{\bsnm{Moody}, \binits{B.E.}},
\bauthor{\bsnm{Lehman}, \binits{L.-W.H.}},
\bauthor{\bsnm{Song}, \binits{C.}},
\bauthor{\bsnm{Li}, \binits{Q.}},
\bauthor{\bsnm{Sun}, \binits{H.}},
\bauthor{\bsnm{Mark}, \binits{R.G.}},
\bauthor{\bsnm{Westover}, \binits{M.B.}},
\bauthor{\bsnm{Clifford}, \binits{G.D.}}:
\bctitle{You snooze, you win: the physionet/computing in cardiology challenge
  2018}.
In: \bbtitle{2018 Computing in Cardiology Conference (CinC)},
vol. \bseriesno{45},
pp. \bfpage{1}--\blpage{4}
(\byear{2018}).
\bcomment{IEEE}
\end{bchapter}
\endbibitem

\bibitem{kemp2000analysis}
\begin{barticle}
\bauthor{\bsnm{Kemp}, \binits{B.}},
\bauthor{\bsnm{Zwinderman}, \binits{A.H.}},
\bauthor{\bsnm{Tuk}, \binits{B.}},
\bauthor{\bsnm{Kamphuisen}, \binits{H.A.}},
\bauthor{\bsnm{Oberye}, \binits{J.J.}}:
\batitle{Analysis of a sleep-dependent neuronal feedback loop: the slow-wave
  microcontinuity of the eeg}.
\bjtitle{IEEE Transactions on Biomedical Engineering}
\bvolume{47}(\bissue{9}),
\bfpage{1185}--\blpage{1194}
(\byear{2000})
\end{barticle}
\endbibitem

\bibitem{quan1997sleep}
\begin{barticle}
\bauthor{\bsnm{Quan}, \binits{S.F.}},
\bauthor{\bsnm{Howard}, \binits{B.V.}},
\bauthor{\bsnm{Iber}, \binits{C.}},
\bauthor{\bsnm{Kiley}, \binits{J.P.}},
\bauthor{\bsnm{Nieto}, \binits{F.J.}},
\bauthor{\bsnm{O'Connor}, \binits{G.T.}},
\bauthor{\bsnm{Rapoport}, \binits{D.M.}},
\bauthor{\bsnm{Redline}, \binits{S.}},
\bauthor{\bsnm{Robbins}, \binits{J.}},
\bauthor{\bsnm{Samet}, \binits{J.M.}}, \betal:
\batitle{The sleep heart health study: design, rationale, and methods}.
\bjtitle{Sleep}
\bvolume{20}(\bissue{12}),
\bfpage{1077}--\blpage{1085}
(\byear{1997})
\end{barticle}
\endbibitem

\bibitem{cummings1990appendicular}
\begin{barticle}
\bauthor{\bsnm{Cummings}, \binits{S.R.}},
\bauthor{\bsnm{Black}, \binits{D.M.}},
\bauthor{\bsnm{Nevitt}, \binits{M.C.}},
\bauthor{\bsnm{Browner}, \binits{W.S.}},
\bauthor{\bsnm{Cauley}, \binits{J.A.}},
\bauthor{\bsnm{Genant}, \binits{H.K.}},
\bauthor{\bsnm{Mascioli}, \binits{S.R.}},
\bauthor{\bsnm{Scott}, \binits{J.C.}},
\bauthor{\bsnm{Seeley}, \binits{D.G.}},
\bauthor{\bsnm{Steiger}, \binits{P.}}, \betal:
\batitle{Appendicular bone density and age predict hip fracture in women}.
\bjtitle{Jama}
\bvolume{263}(\bissue{5}),
\bfpage{665}--\blpage{668}
(\byear{1990})
\end{barticle}
\endbibitem

\bibitem{spira2008sleep}
\begin{barticle}
\bauthor{\bsnm{Spira}, \binits{A.P.}},
\bauthor{\bsnm{Blackwell}, \binits{T.}},
\bauthor{\bsnm{Stone}, \binits{K.L.}},
\bauthor{\bsnm{Redline}, \binits{S.}},
\bauthor{\bsnm{Cauley}, \binits{J.A.}},
\bauthor{\bsnm{Ancoli-Israel}, \binits{S.}},
\bauthor{\bsnm{Yaffe}, \binits{K.}}:
\batitle{Sleep-disordered breathing and cognition in older women}.
\bjtitle{Journal of the American Geriatrics Society}
\bvolume{56}(\bissue{1}),
\bfpage{45}--\blpage{50}
(\byear{2008})
\end{barticle}
\endbibitem

\bibitem{grigg2016visual}
\begin{barticle}
\bauthor{\bsnm{Grigg-Damberger}, \binits{M.M.}}:
\batitle{The visual scoring of sleep in infants 0 to 2 months of age}.
\bjtitle{Journal of clinical sleep medicine}
\bvolume{12}(\bissue{3}),
\bfpage{429}--\blpage{445}
(\byear{2016})
\end{barticle}
\endbibitem

\bibitem{ronneberger2015u}
\begin{bchapter}
\bauthor{\bsnm{Ronneberger}, \binits{O.}},
\bauthor{\bsnm{Fischer}, \binits{P.}},
\bauthor{\bsnm{Brox}, \binits{T.}}:
\bctitle{U-net: Convolutional networks for biomedical image segmentation}.
In: \bbtitle{International Conference on Medical Image Computing and
  Computer-assisted Intervention},
pp. \bfpage{234}--\blpage{241}
(\byear{2015}).
\bcomment{Springer}
\end{bchapter}
\endbibitem

\bibitem{falk2019u}
\begin{barticle}
\bauthor{\bsnm{Falk}, \binits{T.}},
\bauthor{\bsnm{Mai}, \binits{D.}},
\bauthor{\bsnm{Bensch}, \binits{R.}},
\bauthor{\bsnm{{\c{C}}i{\c{c}}ek}, \binits{{\"O}.}},
\bauthor{\bsnm{Abdulkadir}, \binits{A.}},
\bauthor{\bsnm{Marrakchi}, \binits{Y.}},
\bauthor{\bsnm{B{\"o}hm}, \binits{A.}},
\bauthor{\bsnm{Deubner}, \binits{J.}},
\bauthor{\bsnm{J{\"a}ckel}, \binits{Z.}},
\bauthor{\bsnm{Seiwald}, \binits{K.}}, \betal:
\batitle{U-net: deep learning for cell counting, detection, and morphometry}.
\bjtitle{Nature methods}
\bvolume{16}(\bissue{1}),
\bfpage{67}--\blpage{70}
(\byear{2019})
\end{barticle}
\endbibitem

\bibitem{brandt2020unexpectedly}
\begin{barticle}
\bauthor{\bsnm{Brandt}, \binits{M.}},
\bauthor{\bsnm{Tucker}, \binits{C.J.}},
\bauthor{\bsnm{Kariryaa}, \binits{A.}},
\bauthor{\bsnm{Rasmussen}, \binits{K.}},
\bauthor{\bsnm{Abel}, \binits{C.}},
\bauthor{\bsnm{Small}, \binits{J.}},
\bauthor{\bsnm{Chave}, \binits{J.}},
\bauthor{\bsnm{Rasmussen}, \binits{L.V.}},
\bauthor{\bsnm{Hiernaux}, \binits{P.}},
\bauthor{\bsnm{Diouf}, \binits{A.A.}}, \betal:
\batitle{An unexpectedly large count of trees in the west african sahara and
  sahel}.
\bjtitle{Nature}
\bvolume{587}(\bissue{7832}),
\bfpage{78}--\blpage{82}
(\byear{2020})
\end{barticle}
\endbibitem

\bibitem{kingma2014adam}
\begin{botherref}
\oauthor{\bsnm{Kingma}, \binits{D.P.}},
\oauthor{\bsnm{Ba}, \binits{J.}}:
Adam: A method for stochastic optimization.
Preprint at \url{https://arxiv.org/abs/1412.6980}
(2014)
\end{botherref}
\endbibitem

\bibitem{goodfellow2016deep}
\begin{bbook}
\bauthor{\bsnm{Goodfellow}, \binits{I.}},
\bauthor{\bsnm{Bengio}, \binits{Y.}},
\bauthor{\bsnm{Courville}, \binits{A.}}:
\bbtitle{Deep Learning}.
\bpublisher{MIT Press}, \blocation{???}
(\byear{2016})
\end{bbook}
\endbibitem

\bibitem{pan2009survey}
\begin{barticle}
\bauthor{\bsnm{Pan}, \binits{S.J.}},
\bauthor{\bsnm{Yang}, \binits{Q.}}:
\batitle{A survey on transfer learning}.
\bjtitle{IEEE Transactions on knowledge and data engineering}
\bvolume{22}(\bissue{10}),
\bfpage{1345}--\blpage{1359}
(\byear{2009})
\end{barticle}
\endbibitem

\bibitem{mccloskey1989catastrophic}
\begin{botherref}
\oauthor{\bsnm{McCloskey}, \binits{M.}},
\oauthor{\bsnm{Cohen}, \binits{N.J.}}:
Catastrophic interference in connectionist networks: The sequential learning
  problem.
Psychology of Learning and Motivation,
vol. 24,
pp. 109--165.
Academic Press
(1989)
\end{botherref}
\endbibitem

\bibitem{dumoulin2016learned}
\begin{botherref}
\oauthor{\bsnm{Dumoulin}, \binits{V.}},
\oauthor{\bsnm{Shlens}, \binits{J.}},
\oauthor{\bsnm{Kudlur}, \binits{M.}}:
A learned representation for artistic style.
Preprint at \url{https://arxiv.org/abs/1610.07629}
(2014)
\end{botherref}
\endbibitem

\bibitem{de2017modulating}
\begin{botherref}
\oauthor{\bsnm{De~Vries}, \binits{H.}},
\oauthor{\bsnm{Strub}, \binits{F.}},
\oauthor{\bsnm{Mary}, \binits{J.}},
\oauthor{\bsnm{Larochelle}, \binits{H.}},
\oauthor{\bsnm{Pietquin}, \binits{O.}},
\oauthor{\bsnm{Courville}, \binits{A.C.}}:
Modulating early visual processing by language.
Advances in Neural Information Processing Systems
\textbf{30}
(2017)
\end{botherref}
\endbibitem

\bibitem{huang2017arbitrary}
\begin{bchapter}
\bauthor{\bsnm{Huang}, \binits{X.}},
\bauthor{\bsnm{Belongie}, \binits{S.}}:
\bctitle{Arbitrary style transfer in real-time with adaptive instance
  normalization}.
In: \bbtitle{Proceedings of the IEEE International Conference on Computer
  Vision},
pp. \bfpage{1501}--\blpage{1510}
(\byear{2017})
\end{bchapter}
\endbibitem

\bibitem{miyato2018spectral}
\begin{botherref}
\oauthor{\bsnm{Miyato}, \binits{T.}},
\oauthor{\bsnm{Kataoka}, \binits{T.}},
\oauthor{\bsnm{Koyama}, \binits{M.}},
\oauthor{\bsnm{Yoshida}, \binits{Y.}}:
Spectral normalization for generative adversarial networks.
Preprint at \url{https://arxiv.org/abs/1802.05957}
(2018)
\end{botherref}
\endbibitem

\bibitem{xie2020adversarial}
\begin{bchapter}
\bauthor{\bsnm{Xie}, \binits{C.}},
\bauthor{\bsnm{Tan}, \binits{M.}},
\bauthor{\bsnm{Gong}, \binits{B.}},
\bauthor{\bsnm{Wang}, \binits{J.}},
\bauthor{\bsnm{Yuille}, \binits{A.L.}},
\bauthor{\bsnm{Le}, \binits{Q.V.}}:
\bctitle{Adversarial examples improve image recognition}.
In: \bbtitle{Proceedings of the IEEE/CVF Conference on Computer Vision and
  Pattern Recognition},
pp. \bfpage{819}--\blpage{828}
(\byear{2020})
\end{bchapter}
\endbibitem

\bibitem{gong2022sandwich}
\begin{bchapter}
\bauthor{\bsnm{Gong}, \binits{X.}},
\bauthor{\bsnm{Chen}, \binits{W.}},
\bauthor{\bsnm{Chen}, \binits{T.}},
\bauthor{\bsnm{Wang}, \binits{Z.}}:
\bctitle{Sandwich batch normalization: A drop-in replacement for feature
  distribution heterogeneity}.
In: \bbtitle{Proceedings of the IEEE/CVF Winter Conference on Applications of
  Computer Vision},
pp. \bfpage{2494}--\blpage{2504}
(\byear{2022})
\end{bchapter}
\endbibitem

\bibitem{ioffe2015batch}
\begin{bchapter}
\bauthor{\bsnm{Ioffe}, \binits{S.}},
\bauthor{\bsnm{Szegedy}, \binits{C.}}:
\bctitle{Batch normalization: Accelerating deep network training by reducing
  internal covariate shift}.
In: \bbtitle{International Conference on Machine Learning},
pp. \bfpage{448}--\blpage{456}
(\byear{2015}).
\bcomment{PMLR}
\end{bchapter}
\endbibitem

\bibitem{sokolova2009systematic}
\begin{barticle}
\bauthor{\bsnm{Sokolova}, \binits{M.}},
\bauthor{\bsnm{Lapalme}, \binits{G.}}:
\batitle{A systematic analysis of performance measures for classification
  tasks}.
\bjtitle{Information processing \& management}
\bvolume{45}(\bissue{4}),
\bfpage{427}--\blpage{437}
(\byear{2009})
\end{barticle}
\endbibitem

\end{thebibliography}



\begin{thebibliography}{27}
\ifx \bisbn   \undefined \def \bisbn  #1{ISBN #1}\fi
\ifx \binits  \undefined \def \binits#1{#1}\fi
\ifx \bauthor  \undefined \def \bauthor#1{#1}\fi
\ifx \batitle  \undefined \def \batitle#1{#1}\fi
\ifx \bjtitle  \undefined \def \bjtitle#1{#1}\fi
\ifx \bvolume  \undefined \def \bvolume#1{\textbf{#1}}\fi
\ifx \byear  \undefined \def \byear#1{#1}\fi
\ifx \bissue  \undefined \def \bissue#1{#1}\fi
\ifx \bfpage  \undefined \def \bfpage#1{#1}\fi
\ifx \blpage  \undefined \def \blpage #1{#1}\fi
\ifx \burl  \undefined \def \burl#1{\textsf{#1}}\fi
\ifx \doiurl  \undefined \def \doiurl#1{\url{https://doi.org/#1}}\fi
\ifx \betal  \undefined \def \betal{\textit{et al.}}\fi
\ifx \binstitute  \undefined \def \binstitute#1{#1}\fi
\ifx \binstitutionaled  \undefined \def \binstitutionaled#1{#1}\fi
\ifx \bctitle  \undefined \def \bctitle#1{#1}\fi
\ifx \beditor  \undefined \def \beditor#1{#1}\fi
\ifx \bpublisher  \undefined \def \bpublisher#1{#1}\fi
\ifx \bbtitle  \undefined \def \bbtitle#1{#1}\fi
\ifx \bedition  \undefined \def \bedition#1{#1}\fi
\ifx \bseriesno  \undefined \def \bseriesno#1{#1}\fi
\ifx \blocation  \undefined \def \blocation#1{#1}\fi
\ifx \bsertitle  \undefined \def \bsertitle#1{#1}\fi
\ifx \bsnm \undefined \def \bsnm#1{#1}\fi
\ifx \bsuffix \undefined \def \bsuffix#1{#1}\fi
\ifx \bparticle \undefined \def \bparticle#1{#1}\fi
\ifx \barticle \undefined \def \barticle#1{#1}\fi
\bibcommenthead
\ifx \bconfdate \undefined \def \bconfdate #1{#1}\fi
\ifx \botherref \undefined \def \botherref #1{#1}\fi
\ifx \url \undefined \def \url#1{\textsf{#1}}\fi
\ifx \bchapter \undefined \def \bchapter#1{#1}\fi
\ifx \bbook \undefined \def \bbook#1{#1}\fi
\ifx \bcomment \undefined \def \bcomment#1{#1}\fi
\ifx \oauthor \undefined \def \oauthor#1{#1}\fi
\ifx \citeauthoryear \undefined \def \citeauthoryear#1{#1}\fi
\ifx \endbibitem  \undefined \def \endbibitem {}\fi
\ifx \bconflocation  \undefined \def \bconflocation#1{#1}\fi
\ifx \arxivurl  \undefined \def \arxivurl#1{\textsf{#1}}\fi
\csname PreBibitemsHook\endcsname

\bibitem{zhang2018national}
\begin{barticle}
\bauthor{\bsnm{Zhang}, \binits{G.-Q.}},
\bauthor{\bsnm{Cui}, \binits{L.}},
\bauthor{\bsnm{Mueller}, \binits{R.}},
\bauthor{\bsnm{Tao}, \binits{S.}},
\bauthor{\bsnm{Kim}, \binits{M.}},
\bauthor{\bsnm{Rueschman}, \binits{M.}},
\bauthor{\bsnm{Mariani}, \binits{S.}},
\bauthor{\bsnm{Mobley}, \binits{D.}},
\bauthor{\bsnm{Redline}, \binits{S.}}:
\batitle{The national sleep research resource: towards a sleep data commons}.
\bjtitle{Journal of the American Medical Informatics Association}
\bvolume{25}(\bissue{10}),
\bfpage{1351}--\blpage{1358}
(\byear{2018})
\end{barticle}
\endbibitem

\bibitem{bakker2018gastric}
\begin{barticle}
\bauthor{\bsnm{Bakker}, \binits{J.P.}},
\bauthor{\bsnm{Tavakkoli}, \binits{A.}},
\bauthor{\bsnm{Rueschman}, \binits{M.}},
\bauthor{\bsnm{Wang}, \binits{W.}},
\bauthor{\bsnm{Andrews}, \binits{R.}},
\bauthor{\bsnm{Malhotra}, \binits{A.}},
\bauthor{\bsnm{Owens}, \binits{R.L.}},
\bauthor{\bsnm{Anand}, \binits{A.}},
\bauthor{\bsnm{Dudley}, \binits{K.A.}},
\bauthor{\bsnm{Patel}, \binits{S.R.}}:
\batitle{Gastric banding surgery versus continuous positive airway pressure for
  obstructive sleep apnea: a randomized controlled trial}.
\bjtitle{American journal of respiratory and critical care medicine}
\bvolume{197}(\bissue{8}),
\bfpage{1080}--\blpage{1083}
(\byear{2018})
\end{barticle}
\endbibitem

\bibitem{rosen2003prevalence}
\begin{barticle}
\bauthor{\bsnm{Rosen}, \binits{C.L.}},
\bauthor{\bsnm{Larkin}, \binits{E.K.}},
\bauthor{\bsnm{Kirchner}, \binits{H.L.}},
\bauthor{\bsnm{Emancipator}, \binits{J.L.}},
\bauthor{\bsnm{Bivins}, \binits{S.F.}},
\bauthor{\bsnm{Surovec}, \binits{S.A.}},
\bauthor{\bsnm{Martin}, \binits{R.J.}},
\bauthor{\bsnm{Redline}, \binits{S.}}:
\batitle{Prevalence and risk factors for sleep-disordered breathing in 8-to
  11-year-old children: association with race and prematurity}.
\bjtitle{The Journal of pediatrics}
\bvolume{142}(\bissue{4}),
\bfpage{383}--\blpage{389}
(\byear{2003})
\end{barticle}
\endbibitem

\bibitem{redline1995familial}
\begin{barticle}
\bauthor{\bsnm{Redline}, \binits{S.}},
\bauthor{\bsnm{Tishler}, \binits{P.V.}},
\bauthor{\bsnm{Tosteson}, \binits{T.D.}},
\bauthor{\bsnm{Williamson}, \binits{J.}},
\bauthor{\bsnm{Kump}, \binits{K.}},
\bauthor{\bsnm{Browner}, \binits{I.}},
\bauthor{\bsnm{Ferrette}, \binits{V.}},
\bauthor{\bsnm{Krejci}, \binits{P.}}:
\batitle{The familial aggregation of obstructive sleep apnea.}
\bjtitle{American journal of respiratory and critical care medicine}
\bvolume{151}(\bissue{3}),
\bfpage{682}--\blpage{687}
(\byear{1995})
\end{barticle}
\endbibitem

\bibitem{marcus2013randomized}
\begin{barticle}
\bauthor{\bsnm{Marcus}, \binits{C.L.}},
\bauthor{\bsnm{Moore}, \binits{R.H.}},
\bauthor{\bsnm{Rosen}, \binits{C.L.}},
\bauthor{\bsnm{Giordani}, \binits{B.}},
\bauthor{\bsnm{Garetz}, \binits{S.L.}},
\bauthor{\bsnm{Taylor}, \binits{H.G.}},
\bauthor{\bsnm{Mitchell}, \binits{R.B.}},
\bauthor{\bsnm{Amin}, \binits{R.}},
\bauthor{\bsnm{Katz}, \binits{E.S.}},
\bauthor{\bsnm{Arens}, \binits{R.}}, \betal:
\batitle{A randomized trial of adenotonsillectomy for childhood sleep apnea}.
\bjtitle{N Engl J Med}
\bvolume{368},
\bfpage{2366}--\blpage{2376}
(\byear{2013})
\end{barticle}
\endbibitem

\bibitem{redline2011childhood}
\begin{barticle}
\bauthor{\bsnm{Redline}, \binits{S.}},
\bauthor{\bsnm{Amin}, \binits{R.}},
\bauthor{\bsnm{Beebe}, \binits{D.}},
\bauthor{\bsnm{Chervin}, \binits{R.D.}},
\bauthor{\bsnm{Garetz}, \binits{S.L.}},
\bauthor{\bsnm{Giordani}, \binits{B.}},
\bauthor{\bsnm{Marcus}, \binits{C.L.}},
\bauthor{\bsnm{Moore}, \binits{R.H.}},
\bauthor{\bsnm{Rosen}, \binits{C.L.}},
\bauthor{\bsnm{Arens}, \binits{R.}}, \betal:
\batitle{The childhood adenotonsillectomy trial (chat): rationale, design, and
  challenges of a randomized controlled trial evaluating a standard surgical
  procedure in a pediatric population}.
\bjtitle{Sleep}
\bvolume{34}(\bissue{11}),
\bfpage{1509}--\blpage{1517}
(\byear{2011})
\end{barticle}
\endbibitem

\bibitem{perslev2021u}
\begin{barticle}
\bauthor{\bsnm{Perslev}, \binits{M.}},
\bauthor{\bsnm{Darkner}, \binits{S.}},
\bauthor{\bsnm{Kempfner}, \binits{L.}},
\bauthor{\bsnm{Nikolic}, \binits{M.}},
\bauthor{\bsnm{Jennum}, \binits{P.J.}},
\bauthor{\bsnm{Igel}, \binits{C.}}:
\batitle{U-sleep: resilient high-frequency sleep staging}.
\bjtitle{NPJ digital medicine}
\bvolume{4}(\bissue{1}),
\bfpage{1}--\blpage{12}
(\byear{2021})
\end{barticle}
\endbibitem

\bibitem{rosen2012multisite}
\begin{barticle}
\bauthor{\bsnm{Rosen}, \binits{C.L.}},
\bauthor{\bsnm{Auckley}, \binits{D.}},
\bauthor{\bsnm{Benca}, \binits{R.}},
\bauthor{\bsnm{Foldvary-Schaefer}, \binits{N.}},
\bauthor{\bsnm{Iber}, \binits{C.}},
\bauthor{\bsnm{Kapur}, \binits{V.}},
\bauthor{\bsnm{Rueschman}, \binits{M.}},
\bauthor{\bsnm{Zee}, \binits{P.}},
\bauthor{\bsnm{Redline}, \binits{S.}}:
\batitle{A multisite randomized trial of portable sleep studies and positive
  airway pressure autotitration versus laboratory-based polysomnography for the
  diagnosis and treatment of obstructive sleep apnea: the homepap study}.
\bjtitle{Sleep}
\bvolume{35}(\bissue{6}),
\bfpage{757}--\blpage{767}
(\byear{2012})
\end{barticle}
\endbibitem

\bibitem{chen2015racial}
\begin{barticle}
\bauthor{\bsnm{Chen}, \binits{X.}},
\bauthor{\bsnm{Wang}, \binits{R.}},
\bauthor{\bsnm{Zee}, \binits{P.}},
\bauthor{\bsnm{Lutsey}, \binits{P.L.}},
\bauthor{\bsnm{Javaheri}, \binits{S.}},
\bauthor{\bsnm{Alc{\'a}ntara}, \binits{C.}},
\bauthor{\bsnm{Jackson}, \binits{C.L.}},
\bauthor{\bsnm{Williams}, \binits{M.A.}},
\bauthor{\bsnm{Redline}, \binits{S.}}:
\batitle{Racial/ethnic differences in sleep disturbances: the multi-ethnic
  study of atherosclerosis (mesa)}.
\bjtitle{Sleep}
\bvolume{38}(\bissue{6}),
\bfpage{877}--\blpage{888}
(\byear{2015})
\end{barticle}
\endbibitem

\bibitem{blackwell2011associations}
\begin{barticle}
\bauthor{\bsnm{Blackwell}, \binits{T.}},
\bauthor{\bsnm{Yaffe}, \binits{K.}},
\bauthor{\bsnm{Ancoli-Israel}, \binits{S.}},
\bauthor{\bsnm{Redline}, \binits{S.}},
\bauthor{\bsnm{Ensrud}, \binits{K.E.}},
\bauthor{\bsnm{Stefanick}, \binits{M.L.}},
\bauthor{\bsnm{Laffan}, \binits{A.}},
\bauthor{\bsnm{Stone}, \binits{K.L.}},
\bauthor{\bparticle{in} \bsnm{Men Study~Group}, \binits{O.F.}}:
\batitle{Associations between sleep architecture and sleep-disordered breathing
  and cognition in older community-dwelling men: the osteoporotic fractures in
  men sleep study}.
\bjtitle{Journal of the American Geriatrics Society}
\bvolume{59}(\bissue{12}),
\bfpage{2217}--\blpage{2225}
(\byear{2011})
\end{barticle}
\endbibitem

\bibitem{osteoporotic2015relationships}
\begin{botherref}
Relationships between sleep stages and changes in cognitive function in older
  men: the mros sleep study.
Sleep
\textbf{38}(3),
411--421
(2015)
\end{botherref}
\endbibitem

\bibitem{goldberger2000physiobank}
\begin{barticle}
\bauthor{\bsnm{Goldberger}, \binits{A.L.}},
\bauthor{\bsnm{Amaral}, \binits{L.A.}},
\bauthor{\bsnm{Glass}, \binits{L.}},
\bauthor{\bsnm{Hausdorff}, \binits{J.M.}},
\bauthor{\bsnm{Ivanov}, \binits{P.C.}},
\bauthor{\bsnm{Mark}, \binits{R.G.}},
\bauthor{\bsnm{Mietus}, \binits{J.E.}},
\bauthor{\bsnm{Moody}, \binits{G.B.}},
\bauthor{\bsnm{Peng}, \binits{C.-K.}},
\bauthor{\bsnm{Stanley}, \binits{H.E.}}:
\batitle{Physiobank, physiotoolkit, and physionet: components of a new research
  resource for complex physiologic signals}.
\bjtitle{circulation}
\bvolume{101}(\bissue{23}),
\bfpage{215}--\blpage{220}
(\byear{2000})
\end{barticle}
\endbibitem

\bibitem{ghassemi2018you}
\begin{bchapter}
\bauthor{\bsnm{Ghassemi}, \binits{M.M.}},
\bauthor{\bsnm{Moody}, \binits{B.E.}},
\bauthor{\bsnm{Lehman}, \binits{L.-W.H.}},
\bauthor{\bsnm{Song}, \binits{C.}},
\bauthor{\bsnm{Li}, \binits{Q.}},
\bauthor{\bsnm{Sun}, \binits{H.}},
\bauthor{\bsnm{Mark}, \binits{R.G.}},
\bauthor{\bsnm{Westover}, \binits{M.B.}},
\bauthor{\bsnm{Clifford}, \binits{G.D.}}:
\bctitle{You snooze, you win: the physionet/computing in cardiology challenge
  2018}.
In: \bbtitle{2018 Computing in Cardiology Conference (CinC)},
vol. \bseriesno{45},
pp. \bfpage{1}--\blpage{4}
(\byear{2018}).
\bcomment{IEEE}
\end{bchapter}
\endbibitem

\bibitem{kemp2000analysis}
\begin{barticle}
\bauthor{\bsnm{Kemp}, \binits{B.}},
\bauthor{\bsnm{Zwinderman}, \binits{A.H.}},
\bauthor{\bsnm{Tuk}, \binits{B.}},
\bauthor{\bsnm{Kamphuisen}, \binits{H.A.}},
\bauthor{\bsnm{Oberye}, \binits{J.J.}}:
\batitle{Analysis of a sleep-dependent neuronal feedback loop: the slow-wave
  microcontinuity of the eeg}.
\bjtitle{IEEE Transactions on Biomedical Engineering}
\bvolume{47}(\bissue{9}),
\bfpage{1185}--\blpage{1194}
(\byear{2000})
\end{barticle}
\endbibitem

\bibitem{quan1997sleep}
\begin{barticle}
\bauthor{\bsnm{Quan}, \binits{S.F.}},
\bauthor{\bsnm{Howard}, \binits{B.V.}},
\bauthor{\bsnm{Iber}, \binits{C.}},
\bauthor{\bsnm{Kiley}, \binits{J.P.}},
\bauthor{\bsnm{Nieto}, \binits{F.J.}},
\bauthor{\bsnm{O'Connor}, \binits{G.T.}},
\bauthor{\bsnm{Rapoport}, \binits{D.M.}},
\bauthor{\bsnm{Redline}, \binits{S.}},
\bauthor{\bsnm{Robbins}, \binits{J.}},
\bauthor{\bsnm{Samet}, \binits{J.M.}}, \betal:
\batitle{The sleep heart health study: design, rationale, and methods}.
\bjtitle{Sleep}
\bvolume{20}(\bissue{12}),
\bfpage{1077}--\blpage{1085}
(\byear{1997})
\end{barticle}
\endbibitem

\bibitem{cummings1990appendicular}
\begin{barticle}
\bauthor{\bsnm{Cummings}, \binits{S.R.}},
\bauthor{\bsnm{Black}, \binits{D.M.}},
\bauthor{\bsnm{Nevitt}, \binits{M.C.}},
\bauthor{\bsnm{Browner}, \binits{W.S.}},
\bauthor{\bsnm{Cauley}, \binits{J.A.}},
\bauthor{\bsnm{Genant}, \binits{H.K.}},
\bauthor{\bsnm{Mascioli}, \binits{S.R.}},
\bauthor{\bsnm{Scott}, \binits{J.C.}},
\bauthor{\bsnm{Seeley}, \binits{D.G.}},
\bauthor{\bsnm{Steiger}, \binits{P.}}, \betal:
\batitle{Appendicular bone density and age predict hip fracture in women}.
\bjtitle{Jama}
\bvolume{263}(\bissue{5}),
\bfpage{665}--\blpage{668}
(\byear{1990})
\end{barticle}
\endbibitem

\bibitem{spira2008sleep}
\begin{barticle}
\bauthor{\bsnm{Spira}, \binits{A.P.}},
\bauthor{\bsnm{Blackwell}, \binits{T.}},
\bauthor{\bsnm{Stone}, \binits{K.L.}},
\bauthor{\bsnm{Redline}, \binits{S.}},
\bauthor{\bsnm{Cauley}, \binits{J.A.}},
\bauthor{\bsnm{Ancoli-Israel}, \binits{S.}},
\bauthor{\bsnm{Yaffe}, \binits{K.}}:
\batitle{Sleep-disordered breathing and cognition in older women}.
\bjtitle{Journal of the American Geriatrics Society}
\bvolume{56}(\bissue{1}),
\bfpage{45}--\blpage{50}
(\byear{2008})
\end{barticle}
\endbibitem

\bibitem{Ohayon2004}
\begin{barticle}
\bauthor{\bsnm{Ohayon}, \binits{M.}},
\bauthor{\bsnm{Carskadon}, \binits{M.}},
\bauthor{\bsnm{Guilleminault}, \binits{C.}},
\bauthor{\bsnm{Vitiello}, \binits{M.}}:
\batitle{Meta-analysis of quantitative sleep parameters from childhood to old
  age in healthy individuals: Developing normative sleep values across the
  human lifespan}.
\bjtitle{Sleep}
\bvolume{27},
\bfpage{1255}--\blpage{73}
(\byear{2004})
\end{barticle}
\endbibitem

\bibitem{berry2012rules}
\begin{barticle}
\bauthor{\bsnm{Berry}, \binits{R.B.}},
\bauthor{\bsnm{Budhiraja}, \binits{R.}},
\bauthor{\bsnm{Gottlieb}, \binits{D.J.}},
\bauthor{\bsnm{Gozal}, \binits{D.}},
\bauthor{\bsnm{Iber}, \binits{C.}},
\bauthor{\bsnm{Kapur}, \binits{V.K.}},
\bauthor{\bsnm{Marcus}, \binits{C.L.}},
\bauthor{\bsnm{Mehra}, \binits{R.}},
\bauthor{\bsnm{Parthasarathy}, \binits{S.}},
\bauthor{\bsnm{Quan}, \binits{S.F.}}, \betal:
\batitle{Rules for scoring respiratory events in sleep: update of the 2007 aasm
  manual for the scoring of sleep and associated events: deliberations of the
  sleep apnea definitions task force of the american academy of sleep
  medicine}.
\bjtitle{Journal of clinical sleep medicine}
\bvolume{8}(\bissue{5}),
\bfpage{597}--\blpage{619}
(\byear{2012})
\end{barticle}
\endbibitem

\bibitem{danker2009interrater}
\begin{barticle}
\bauthor{\bsnm{Danker-hopfe}, \binits{H.}},
\bauthor{\bsnm{Anderer}, \binits{P.}},
\bauthor{\bsnm{Zeitlhofer}, \binits{J.}},
\bauthor{\bsnm{Boeck}, \binits{M.}},
\bauthor{\bsnm{Dorn}, \binits{H.}},
\bauthor{\bsnm{Gruber}, \binits{G.}},
\bauthor{\bsnm{Heller}, \binits{E.}},
\bauthor{\bsnm{Loretz}, \binits{E.}},
\bauthor{\bsnm{Moser}, \binits{D.}},
\bauthor{\bsnm{Parapatics}, \binits{S.}}, \betal:
\batitle{Interrater reliability for sleep scoring according to the
  rechtschaffen \& kales and the new aasm standard}.
\bjtitle{Journal of sleep research}
\bvolume{18}(\bissue{1}),
\bfpage{74}--\blpage{84}
(\byear{2009})
\end{barticle}
\endbibitem

\bibitem{rosenberg2013american}
\begin{barticle}
\bauthor{\bsnm{Rosenberg}, \binits{R.S.}},
\bauthor{\bsnm{Van~Hout}, \binits{S.}}:
\batitle{The american academy of sleep medicine inter-scorer reliability
  program: sleep stage scoring}.
\bjtitle{Journal of clinical sleep medicine}
\bvolume{9}(\bissue{01}),
\bfpage{81}--\blpage{87}
(\byear{2013})
\end{barticle}
\endbibitem

\bibitem{younes2016staging}
\begin{barticle}
\bauthor{\bsnm{Younes}, \binits{M.}},
\bauthor{\bsnm{Raneri}, \binits{J.}},
\bauthor{\bsnm{Hanly}, \binits{P.}}:
\batitle{Staging sleep in polysomnograms: analysis of inter-scorer
  variability}.
\bjtitle{Journal of Clinical Sleep Medicine}
\bvolume{12}(\bissue{06}),
\bfpage{885}--\blpage{894}
(\byear{2016})
\end{barticle}
\endbibitem

\bibitem{muto20180315}
\begin{barticle}
\bauthor{\bsnm{Muto}, \binits{V.}},
\bauthor{\bsnm{Berthomier}, \binits{C.}},
\bauthor{\bsnm{Schmidt}, \binits{C.}},
\bauthor{\bsnm{Vandewalle}, \binits{G.}},
\bauthor{\bsnm{Jaspar}, \binits{M.}},
\bauthor{\bsnm{Devillers}, \binits{J.}},
\bauthor{\bsnm{Chellappa}, \binits{S.}},
\bauthor{\bsnm{Meyer}, \binits{C.}},
\bauthor{\bsnm{Phillips}, \binits{C.}},
\bauthor{\bsnm{Berthomier}, \binits{P.}}, \betal:
\batitle{0315 inter-and intra-expert variability in sleep scoring: comparison
  between visual and automatic analysis}.
\bjtitle{Sleep}
\bvolume{41}(\bissue{suppl\_1}),
\bfpage{121}
(\byear{2018})
\end{barticle}
\endbibitem

\bibitem{goodfellow2016deep}
\begin{bbook}
\bauthor{\bsnm{Goodfellow}, \binits{I.}},
\bauthor{\bsnm{Bengio}, \binits{Y.}},
\bauthor{\bsnm{Courville}, \binits{A.}}:
\bbtitle{Deep Learning}.
\bpublisher{MIT Press}, \blocation{???}
(\byear{2016})
\end{bbook}
\endbibitem

\bibitem{lukasik2020does}
\begin{bchapter}
\bauthor{\bsnm{Lukasik}, \binits{M.}},
\bauthor{\bsnm{Bhojanapalli}, \binits{S.}},
\bauthor{\bsnm{Menon}, \binits{A.}},
\bauthor{\bsnm{Kumar}, \binits{S.}}:
\bctitle{Does label smoothing mitigate label noise?}
In: \bbtitle{International Conference on Machine Learning},
pp. \bfpage{6448}--\blpage{6458}
(\byear{2020}).
\bcomment{PMLR}
\end{bchapter}
\endbibitem

\bibitem{fiorillo2021deepsleepnet}
\begin{barticle}
\bauthor{\bsnm{Fiorillo}, \binits{L.}},
\bauthor{\bsnm{Favaro}, \binits{P.}},
\bauthor{\bsnm{Faraci}, \binits{F.D.}}:
\batitle{Deepsleepnet-lite: A simplified automatic sleep stage scoring model
  with uncertainty estimates}.
\bjtitle{IEEE Transactions on Neural Systems and Rehabilitation Engineering}
\bvolume{29},
\bfpage{2076}--\blpage{2085}
(\byear{2021})
\end{barticle}
\endbibitem

\bibitem{szegedy2016rethinking}
\begin{bchapter}
\bauthor{\bsnm{Szegedy}, \binits{C.}},
\bauthor{\bsnm{Vanhoucke}, \binits{V.}},
\bauthor{\bsnm{Ioffe}, \binits{S.}},
\bauthor{\bsnm{Shlens}, \binits{J.}},
\bauthor{\bsnm{Wojna}, \binits{Z.}}:
\bctitle{Rethinking the inception architecture for computer vision}.
In: \bbtitle{Proceedings of the IEEE Conference on Computer Vision and Pattern
  Recognition},
pp. \bfpage{2818}--\blpage{2826}
(\byear{2016})
\end{bchapter}
\endbibitem

\end{thebibliography}

\newpage

{

\section*{Figure legends}

\textbf{Figure~\ref{fig:usleep_architecture}. U-Sleep overall architecture}. U-Sleep is a fully convolutional deep neural network. It takes as input a sequence of length $L$ of 30-second sleep epochs and it outputs the predicted sleep stage for each epoch. We slightly modified the original figure (see \textit{Figure 2: Model architecture} in \cite{perslev2021u}) reporting the additional SaBN layers exploited in the conditional learning procedure (see subsection Conditional learning). Please refer to \cite{perslev2021u} for details on the U-Sleep model architecture and training parameters.\\

\section*{Table legends}

\textbf{Table~\ref{db_overview}. Datasets overview with demographic statistics}. Missing values are due to study design or anonymized data. On the BSDB dataset, we compute the age and the sex values on the 99.1\% and on the 98.6\% of the whole dataset, respectively, because of missing age/sex information. Datasets directly available online are identified by \checkmark, whilst datasets that require approval from a Data Access Committee marked by (\checkmark). BSDB is a private dataset.\\

\textbf{Table~\ref{usleep_i}. \textit{(i)} Clinically non-recommended channel derivations}. Performance of U-Sleep-v0 and U-Sleep-v1, pre-trained on the OA datasets, and evaluated on the test set of the BSDB dataset {(data split in Supplementary Table 8)}, and on the whole BSDB\textsubscript{(100\%)} dataset, \textit{i.e.}, both direct transfer (DT) on BSDB. We report the F1-score (\%F1), specifically the mean value and the standard deviation $(\mu \pm \sigma)$ computed across the recordings.\\

\textbf{Table~\ref{usleep_ii}. \textit{(ii)} Generalizability on different data centers with a heterogeneous dataset}. Performance of U-Sleep-v1, pre-trained on the OA datasets, and evaluated on all the test sets of the OA datasets and on the test set of the BSDB dataset {(data split in Supplementary Table 7 and Supplementary Table 8)}. We also report the performance of U-Sleep-v1 trained from scratch (S) or fine-tuned (FT) on the BSDB dataset, and evaluated on all the test sets of all the available datasets. We report the F1-score (\%F1), specifically the mean value and the standard deviation $(\mu \pm \sigma)$ computed across the recordings.\\

\textbf{Table~\ref{usleep_iii}. \textit{(iii)} Training conditioned by age}. Performance of U-Sleep-v1 on a single model fine-tuned on all the training set of the seven BSDB groups (FT-G1); on seven/two models fine-tuned on the independent training set of each group with $G=7$ (FT-I-G7) and $G=2$ (FT-I-G2) respectively; and on a single model fine-tuned on all the training set of the seven/two BSDB groups conditioned by $G=7$ (FT-SaBN-G7) and by $G=2$ (FT-SaBN-G2) groups respectively. All the fine-tuned models are evaluated on the associated test set of each group {(data split in Supplementary Table 8)}. We report the F1-score (\%F1), specifically the mean value and the standard deviation $(\mu \pm \sigma)$ computed across the recordings. B: Babies (0-3 years); C: Children (4-12 years); A: Adolescents (13-18 years); YA: Young Adults (19-39 years); MA: Middle Aged Adults (40-59 years); E: Elderly (60-69 years); OE: Old Elderly ($\ge$ 70 years). When $G=2$ we have the following two groups $G\textsubscript{1}=\{B\cup C\}$, $G\textsubscript{2}=\{A\cup YA\cup MA\cup E\cup OE\}$, further details in Supplementary notes: Age analysis.

}

\end{document}